\documentclass[12pt, oneside]{article}   
\usepackage{amssymb,amsmath,amsthm,enumerate,hyperref,mathtools}    
\usepackage[dvipsnames]{xcolor}    
\usepackage{ulem}   
\hypersetup{
    colorlinks, 
    linkcolor={red!50!black}, 
    citecolor={blue!50!black}, 
    urlcolor={blue!80!black}  
}
\usepackage[displaymath, mathlines]{lineno}  % elsevier
\usepackage{mathabx}  
\usepackage{graphicx,latexsym,amsfonts}  
\usepackage{mathrsfs}  
\usepackage{pstricks}
\usepackage{lineno}
\usepackage[english]{babel}
\usepackage{diagbox}
\usepackage{slashbox} 
\usepackage{pict2e}
\usepackage{picture}

\renewcommand{\emph}{\textsl}

\def\reference#1{\href{#1}{Cliquer ici pour voir une r\'ef\'erence.}} 

\usepackage[longnamesfirst]{natbib}
\definecolor{aquamarine}{rgb}{0, 0.68, 0.62}

\author{
Davide Carpentiere\thanks{\scriptsize Department of Mathematics and Computer Science, University of Catania, Italy.},
Alfio Giarlotta\thanks{\scriptsize Department of Economics and Business, University of Catania, Italy. alfio.giarlotta@unict.it \textit{(corresponding author)}}$\:$,
Stephen Watson\thanks{\scriptsize Department of Mathematics and Statistics, York University, Toronto, Canada.}}

\def\R{{\mathbb R}}

\def\Choice{\textsf{Choice}}
\def\choice{\textsf{choice}}
\def\LO{\textsf{LO}}

\def\XX{\mathscr{X}} 
\def\es{\varnothing}

\def\irr{{\operatorname{irr}}} 
\def\rat{{\operatorname{rat}}}

\def\Par{{\operatorname{Par}}}
\def\qed{\hfill $\Box$ \\}

\def\vs{\vspace{-0,1cm}}

\newtheorem{theorem}{Theorem}[section]

\newtheorem{lemma}[theorem]{Lemma}
\newtheorem{remark}[theorem]{Remark}
\newtheorem{example}[theorem]{Example}

\newtheorem{definition}[theorem]{Definition}

\title{Rational measures of irrationality}

\date{}

\begin{document}

\maketitle

\vspace{-0.5cm}

\begin{abstract}
\noindent   
All possible types of deterministic choice behavior are classified by their degree of irrationality. 
This classification is performed in three steps: (1) select a benchmark of rationality, for which this degree is zero; (2) endow the set of choices with a metric to measure deviations from rationality; and (3) compute the distance of any choice behavior from the selected benchmark. 
The natural candidate for step 1 is the family of all rationalizable behaviors. %according to revealed preference theory. 
A possible candidate for step 2 is a suitable variation of the metric described by \citet{Klamler2008}, which displays a sharp discerning power among different types of choice behaviors. 
%A quasi-choice over a finite set $X$ is a function that assigns to any subset $A$ of $X$ a possibly empty subset of $A$. 
%However, his claim does not hold. 
%However, Klamler's main result fails to hold. 
%While proving a correct characterization of this metric, we determine the causes of its low discriminating power, and design a high-discerning variation of it. % sharpen it using a notion of local rationalization.  %to assign fine levels of irrationality to choice behaviors. 
%Theorem~1 of \cite{Klamler2008} says that the symmetric difference on sets induces a  natural metric on quasi-choices, and this is the unique distance satisfying some desirable properties.
%Unfortunately, Klamler's result is incorrect, because his set of axioms guarantees neither existence nor uniqueness. 
%We amend two axioms, and prove a characterization.  
%However, Klamler's metric can hardly be used to assign a degree of rationality of any quasi-choice, because its discernibility power is quite unrefined. 
%This allows us to sharpen of Klamler's metric by a notion of local rationalization, which displays a high discerning power of different behaviors. 
In step 3 we use this new metric to establish the minimum distance of any choice behavior from the benchmark of rationality.
%This measure is performed by computing the minimum distance from the benchmark.  
Finally we describe a measure of stochastic irrationality, which employs the random utility model as a benchmark of rationality, and the Block-Marschak polynomials to measure deviations from it. 
% encoding the subjective desirability of rational behavior by its degree of transitivity, and then deriving a weighted measure of irrationality. 
\medskip


\medskip

\noindent \textbf{\textsf{JEL} Classification:} D01, D81, C44.

\end{abstract}

%%%%%%%%%%%%%%%%%%%%%%%%%%%%%%
%%%%%%%%%%%%%%%%%%%%%%%%%%%%%%
%%%%%%%%%%%%%%%%%%%%%%%%%%%%%%
%%%%%%%%%%%%%%%%%%%%%%%%%%%%%%

\section{Introduction} \label{SECT:intro}

The goal of this paper is to evaluate the irrationality level of all possible choice behaviors on a finite set of alternatives. 
We perform this task in three successive steps:
\begin{itemize}
	\item[(1)] establish a benchmark of rational choice behavior;
	\item[(2)] endow the set of all choice behaviors with a highly discerning metric;
	\item[(3)] compute the distance of any behavior from the benchmark of rationality.
\end{itemize}
The output of this process is a \textsl{rational degree of irrationality} of any deterministic choice behavior.
(The use of the term `rational' is motivated by the fact that we compute a distance from rationality in order to measure irrationality.)   
Before addressing in detail each step of this approach, let us discuss the general domain of our analysis. %that is, the type of choice behavior under examination.  
%
%To make the scope of our analysis as general as possible, we also consider indecisive choice behaviors, that is, \textsl{quasi-choices}: from any subset of the given set of alternatives, the decision maker (DM) is allowed to select any number of items, one, many, or none. 

Classically, the literature on choice theory is exclusively concentrated on `decisive' choice behaviors, intended as situations in which the decision maker (DM) selects at least one item from any nonempty subset of the ground set: see, among a large amount of relevant contributions, the seminal papers by \cite{Samuelson1938}, \cite{Arrow1959}, and \cite{Sen1971}. 
In other words, the domain of analysis is classically restricted to \textsl{choice correspondences}, which are functions mapping nonempty sets into nonempty subsets.   
In addition, most of the recent models of `bounded rationality in choice' typically deal with the even more restricted case of \textsl{choice functions}, which are single-valued choices correspondences (i.e., a unique item is selected from any nonempty menu): see, among several papers on the topic, \cite{ManziniMariotti2007} and \cite{MasatliogluNakajimaOzbay2012}.\footnote{See \cite{GiaPetWat2022a} for a list of many models of bounded rationality in choice and a common analysis of their features by a unified approach.} 

Despite the great abundance of literature on choice functions and choice correspondences, it appears more realistic to consider the general case of \textsl{quasi-choices}, which model the behavior of possibly indecisive DMs: in this situation, the agent is allowed to select all, some, or none of the items available in any menu. 
To justify the potential interest in this approach, very recently \cite{Costa-GomezCuevaGerasimouTejiscak2022} mention some compelling experiments, which suggest that choice models rejecting decisiveness may offer a powerful tool to study revealed preferences.\footnote{See also Chapter~1 of the advanced textbook on microeconomic theory by \citet{Kreps2013}, as well as the arguments presented in Section~1 of the recent paper by \citet{AlcantudCantoneGiarlottaWatson2022}.}  
That is why in this paper we evaluate the rationality level of any type of choice behavior, may it be decisive or not.  

Now we describe the three stages of our approach. 

\begin{description}
	\item[(1)] The first step consists of the selection of the benchmark of rationality ---the `zero'--- from which deviations ought  to be measured. 
	We select the most natural candidate, namely the family of all quasi-choices over the given set that are considered `rational'  according to \textsl{revealed preference theory} \citep{Samuelson1938}.
	Technically, these are the quasi-choices that can be explained by the maximization of a binary relation.\footnote{A more restrictive benchmark of rationality may be the family of quasi-choices rationalizable by binary relations satisfying some desirable properties.}  
%	In this setting, all quasi-choices belonging to the benchmark of rationality will have an irrationality index equal to zero, whereas all the others will have an index with positive value.   
%
	\item[(2)] The selection of a metric is the key step: this distance should accurately discerns among different types of choice behavior in an economically significant way.   
%	In fact, in order to consistently compute deviations from the benchmark of rationality, here we select a metric that accurately discerns among different types of choice behavior in an economically significant way. 
	A possible candidate for this goal is the distance on quasi-choices proposed by \cite{Klamler2008}, which is computed by summing the cardinalities of all symmetric differences between pairs of choice sets. 
%	The author proves that this metric satisfies five intuitive properties, and is the unique distance that does so.  
%	Regrettably, its characterization is incorrect, because neither existence nor uniqueness is guaranteed. 
%	We amend two of these five properties, and obtain a correct characterization.  
	Due to its decomposability into trivial metrics, Klamler's distance is however not well-suited for our goals, due to its low discernibility power.  
	Using a notion of local rationalization, we design a refinement of this metric, which displays a sharp level of discrimination among different choices. 
	\item[(3)] To finally establish the degree of irrationality for any deterministic choice behavior, we use the metric selected at step 2 to compute the minimum distance of a quasi-choice from a rationalizable one.   
	In this way, all quasi-choices belonging to the benchmark of rationality have a degree of irrationality equal to zero, whereas all the others display a degree with a strictly positive value.   	
	Moreover, the more irrational a choice behavior is, the higher the value of the index becomes. 
	We also describe a weighted version of this approach.  
	Formally, since each rationalizable choice is explained by the maximization of a unique asymmetric preference ---the \textsl{strict revealed preference} \citep{Samuelson1938}---, we measure the subjective desirability of each rational behavior by the `level of transitivity' of this binary relation:\footnote{Both \citet{Mas-ColellWhinstonGreen1995} and \cite{Kreps2013} consider transitivity and completeness the basic tenets of economic rationality.} the more this preference is close to being fully transitive,\footnote{By `fully transitive' we mean that both strict preference and associated incomparability are transitive.} the higher the desirability of the choice becomes. 
%From this point of view, the most desirable behaviors are those rationalized by a \textsl{weak order} (asymmetric and negatively transitive, hence transitive). 
%Slightly less desirable levels are assigned to choices rationalized by \textsl{semiorders} \citep{Luce1956} and \textsl{interval orders} \citep{Fishburn1970,Fishburn1985}. 
%At an even lower desirability level lie all transitive asymmetric relations that fail to be interval orders. 
%At the bottom of the scale, we find those choices that are rationalized by acyclic and intransitive binary relations.\footnote{More generally, the level of transitivity of any acyclic binary relation can be measured by the satisfaction of several types of \textsl{strict and weak $(m,n)$-Ferrers properties} \citep{GiaWat2014Ferrers,GiaWat2018}. On the point, see also \cite{CanGiaGreWat2016} for a classification of all rationalizable choices on the basis of the so-called \textsl{axioms of $(m,n)$-replacement consistency}.}  
Once subjective desirability is encoded, we measure the degree of irrationality of any behavior by taking a weighted distance from rational behavior. 	
\end{description}

Finally, we suggest a probabilistic extension of our approach, which applies to stochastic choice functions.
Recall that a stochastic choice function assigns a real number to each pair formed by a menu and an item in it, evaluating the likelihood of that item being selected from that menu.
Choice functions are special stochastic choices in which this likelihood is one for exactly one item in a menu and zero for all the others. 
%that subset, such that every restriction to every powerset is a probability distribution.
%Specifically, this function encodes the likelihood of selecting an item in a menu in such a way that, when the subset is fixed, the sum of the probability over all items amounts to $1$.
%\textbf{\davide (I made some changes in what follows.)}

The steps to measure the irrationality of a stochastic choice behavior are, however, different from the ones of the  deterministic setting. 
Specifically, the first step is again the selection of a benchmark of rationality, for which we take the family of all stochastic choice functions satisfying the \textsl{random utility model} \textsl{(RUM)} \citep{Block_Marschak1960}. 
On the other hand, since the second and the third step of the deterministic approach are hardly adaptable,  
%cannot be adapted to a stochastic setting. 
%There are two main reasons for this impossibility.   
%First, all metrics ---or weaker variations of them, called \textsl{divergences}--- present in the literature fail to be sufficiently discerning for our purposes, and it looks rather difficult to design a new, highly discerning metric/divergence.  
%Second, and more important, computing the minimum distance from the benchmark of rationality is computationally unfeasible, because there are infinitely may RUM choices.   
%
we employ a different procedure.  
In fact, we take advantage of the characterization established by \cite{Falmagne1978}, who shows that a stochastic choice function satisfies RUM if and only if all its \textsl{Block-Marschak polynomials} are non-negative.
Therefore, any choice that fails to satisfy RUM must have at least one negative Block-Marschak polynomial. 
Upon summing up all these negative polynomials for each element in the ground set, we obtain a \textsl{negativity vector}, which provides a discerning measure of the irrationality of a stochastic choice behavior.  
 The comparison of these vectors is then performed by a permutation-invariant Pareto ordering, which in turn yields a partial classification of all stochastic choices according to their degree of irrationality. 

% 
%\bigskip\bigskip
%
%\textbf{\blue Transfer/eliminate what is below.}
%
%Let $X$ be a finite set of alternatives. 
%A quasi-choice over $X$ is a map $C$ that assigns to each subset $A$ of $X$ a possibly empty subset $C(A)$ of $A$, which collects all selectable elements of $A$. 
%In 2008, Klamler introduced a metric $d_\Delta$ on the family of all quasi-choices over a given set.\footnote{\cite{Klamler2008} denotes this metric by $d_F$. We prefer to use $d_\Delta$ instead, because this notation explicitly indicates the genesis of the metric.} 
%The metric $d_\Delta$ is based on the symmetric difference $\Delta$ of sets \citep{Kemeny1959}. %which evaluates the distance between two sets by only counting the elements contained in exactly one of the two sets.
%Specifically, \cite{Klamler2008} defines the distance between two quasi-choices over $X$ by summing up all cardinalities of the symmetric differences between choice sets. 
%His main result (Theorem~1) states that $d_\Delta$ is the unique metric that satisfies five (desirable) properties \textsf{A1}--\textsf{A5}.
 \smallskip

The paper is organized as follows. 
Section~\ref{SECT:measuring_irrationality} collects preliminary notions and presents a review of the related literature on deterministic choices. 
In Section~\ref{SECT:metrics_on_choices} we describe the metric introduced by Klamler, and then a highly discerning variation of it. %detect some issue in the known characterization, and prove a correct characterization. 
%In Section~\ref{SECT:rational_metric} we propose a sharpening of Klamler's metric based on a notion of local rationalization.    
In Section~\ref{SECT:rational_index} we formally define two distance-based degrees of irrationality of a choice behavior, and show the soundness of the novel metric for this task.  
In Section~\ref{SECT:stochastic_setting} we suggest an extension of our approach to a stochastic environment.  

%%%%%%%%%%%%%%%%%%%%%%%%%%%%%%
%%%%%%%%%%%%%%%%%%%%%%%%%%%%%%
%%%%%%%%%%%%%%%%%%%%%%%%%%%%%%
%%%%%%%%%%%%%%%%%%%%%%%%%%%%%%

\section{Measures of deterministic irrationality} \label{SECT:measuring_irrationality}

First we recall some preliminary notions in choice theory. 
Then we suggest several ways to measures the irrationality of a deterministic choice behavior, and present a quick review of recent literature on the topic.  
%Successively, we introduce the metric based on the symmetric difference of choice sets, and state the characterization of this metric given by \citet[Theorem~1]{Klamler2008}.   

%%%%%%%%%%%%%%%%%%%%%%%%%%%%%%
%%%%%%%%%%%%%%%%%%%%%%%%%%%%%%

\subsection{Preliminaries}

%Here we recall the classical notion of rationalizability. 
A finite set $X$ of $n \geq 2$ alternatives is fixed throughout.
We use $\XX$ to denote the family of all nonempty subsets of $X$.
%We recall the notions of a quasi-choice correspondence and a choice correspondence.

A \textsl{quasi-choice correspondence} over $X$ is a function $C \colon \XX \cup \{\es\} \to \XX \cup \{\es\}$ such that $C(A) \subseteq A$ for all $A \in \XX \cup \{\es\}$. %\footnote{Note that it suffices to define a quasi-choice on the family $\XX$ of nonempty subsets of $X$. Therefore, to avoid a cumbersome notation, sometimes we use $\XX$ in place of $\XX \cup \{\es\}$.} 
A \textsl{choice correspondence} over $X$ is a quasi-choice that is never empty-valued on nonempty sets, that is, a function $c \colon \XX \cup \{\es\} \to \XX \cup \{\es\}$ such that $\varnothing \neq c(A) \subseteq A$ for all $A \in \XX$.\footnote{To emphasize decisiveness, we shall use upper case letters ($C$, $C'$, etc.) to denote possibly indecisive choice behaviors, that is, quasi-choice correspondences. On the other hand, lower case letters ($c$, $c'$, etc.) will be employed to denote decisive choice behaviors, that is, choice correspondences.} 
Sets in $\XX \cup \{\es\}$ are \textsl{menus}, elements of a nonempty menu are \textsl{items}, and the set $C(A)$ (or $c(A)$) is the \textsl{choice set} of the menu $A$.
Unless confusion may arise, hereafter we speak of \textsl{quasi-choices} and \textsl{choices}, respectively.   
Moreover, $\Choice(X)$ (resp.\ $\textsf{choice}(X)$) denotes the family of all quasi-choices (resp.\ choices) over $X$.
\smallskip

A binary relation $\succ$ over $X$ is a subset of $X \times X$, which is:

-  \textsl{asymmetric} if $x \succ y$ implies $\neg(y \succ x)$ for all $x,y \in X$;

- \textsl{irreflexive} if $x \succ x$ holds for no $x \in X$;

- \textsl{acyclic} if $x_1 \succ x_2 \succ \ldots \succ x_k \succ x_1$ holds for no $x_1,x_2,\ldots,x_k \in X$ ($k \geqslant 3$);

- \textsl{transitive} if $x \succ y \succ z$ implies $x \succ z$ for all $x,y,z \in X$;

- \textsl{negatively transitive} if $\neg(x \succ y) \wedge \neg(y \succ z)$ implies $\neg(x \succ z)$ for all $x,y,z \in X$.

\noindent Note that (i) asymmetry implies irreflexivity, (ii) transitivity and asymmetry implies acyclicity, and (iii) asymmetry and negative transitivity implies transitivity.  
We will often refer to an asymmetric binary relation as a \textsl{(strict) preference}.
\smallskip

Choices and preferences are closely related to each other. 
In fact, since the seminal work of \cite{Samuelson1938}, the `rationality' of a decisive choice behavior is classically modeled by the notion of `binary rationalizability', that is, the possibility to explain it by maximizing a suitable binary relation. 
Formally, a choice $c \colon \XX \cup \{\es\}\to \XX \cup \{\es\}$ is \textsl{rationalizable} if there is an asymmetric binary relation $\succ$ over $X$ such that for any nonempty menu $A$, the equality\vs 
$$
c(A) = \max (A, \succ) = \{x \in A : a \succ x \text{ for no } a \in A\}\vs
$$
holds.
The binary relation $\succ$ is called the \textsl{(strict) preference revealed by $c$}. 
Note that $\succ$ must also be acyclic in order to rationalize the choice $c$. 
Moreover, the asymmetric relation of revealed preference is unique for any rationalizable choice.\footnote{Here, we purposely avoid mentioning the symmetric part of the relation of reveled preference, because it is irrelevant to detect the rationalizability of a choice.}  

%%%%%%%%%%%%%%%%%%%%%%%%%%%%%%
%%%%%%%%%%%%%%%%%%%%%%%%%%%%%%

\subsection{Related literature}

In view of our goal to distinguish choice behaviors by their consistency features, the notion of rationalizability is the most popular in the literature. 
This notion was first introduced for \textsl{choice functions} (that is, single-valued choice correspondences), and then extended to choice correspondences.
However, rationalizability can be naturally generalized to quasi-choices, provided that the rationalizing preference is allowed not to be irreflexive, asymmetric, or acyclic. 
Formally, we call a quasi-choice $C \colon \XX \cup \{\es\} \to \XX \cup \{\es\}$ \textsl{rationalizable} if there is an arbitrary binary relation over $X$ ---here denoted by `$\to$' to emphasize its arbitrariness--- such that the equality\vs 
$$
C(A) = \max(A,\to) = \{x \in A : a \to x \text{ for no } a \in A\}\vs
$$ 
holds for all menus $A \in \XX$.   
Here the key fact is the possible lack of properties of $\to$, which follows from the necessity to model indecisive choice behaviors.   
For instance, since asymmetry is not guaranteed, we may have $x \to y \to x$ for some distinct elements $x$ and $y$, in which case $C(\{x,y\})$ is empty.\footnote{To justify such a situation, imagine a political ballot in which the two remaining candidates are extremists, and my moderate political view suggests me to abstain from voting.}  
Similarly, the possible lack of irreflexivity of $\to$ permits situations of the type $x \to x$, which in turn yields $C(\{x\}) = \es$.\footnote{For instance, if a restaurant only offers a chocolate cake as dessert and I am allergic to chocolate, then I shall avoid taking dessert.}
Note also that, contrary to the case of choices, the rationalizable preference ---which is called a \textsl{voter} by \cite{AlcantudCantoneGiarlottaWatson2022}--- need not be unique for the general case of quasi-choices.\footnote{On this point, see Section~2 in \cite{AlcantudCantoneGiarlottaWatson2022}. Here the authors extensively dwell on the reasons motivating the more general use of quasi-choices instead of choices, and the use of arbitrary binary relations to justify choice behavior.} 

All in all, according to this classical paradigm, any (decisive or indecisive) choice behavior is regarded  \textsl{irrational} if it fails to be rationalizable. 
This yields a simple dichotomy \textit{rational/irrational} or, equivalently, \textit{rationalizable/non-rationalizable}.  
However, this dichotomy not very satisfactory in practice, because rationalizability fails to explain the overwhelming  majority of observed choice behaviors.\footnote{For a precise computation of the fraction of rationalizable choices over a set of fixed size, see \citet[Lemma~6]{GiaPetWat2022a} } 

Recently, following the inspiring analysis of~\cite{Simon1955}, the notion of rationalizability has been amended by several forms of \textsl{bounded rationality}, which aim to explain a larger portion of choice behaviors by means of more flexible paradigms.  
To wit this trend, there are tens of models of bounded rationality in choice that have been proposed in the last twenty years: see \cite{GiaPetWat2022a} for a vast account of them. 
The dichotomy \textit{boundedly rational/boundedly irrational} is certainly more satisfactory than the rational/irrational one, allowing one to identify choice behaviors that obey some more relaxed (but still justifiable) constraints.\footnote{The fraction of boundedly rational choice functions is definitively larger than that of rationalizable choices: compare Lemma~6 with Theorem~3 in \cite{GiaPetWat2022a}.}  
However, this bounded rationality approach does not apply to most choice behaviors: in fact, it has essentially been proposed exclusively for choice functions, with very few cases of choice correspondences, moreover leaving completely out the case of quasi-choices. 

A conceptually different modelization of rationality does not distinguish between (bounded) rationality and (bounded) irrationality.   
Rather, it creates a partition of the family of choices in several classes, each of which is assigned a degree of rationality. 
A seminal approach in this direction is the \textsl{rationalization by multiple rationales (RMR)} of \cite{KalaiRubinsteinSpiegler2002}.
The RMR model yields a partition of the family of all choice functions over a set with $n$ items into $n-1$ equivalence classes of rationality, which are determined by the minimum number of linear orders that are necessary to explain decisive choice behavior: the larger this number, the less rational the behavior.\footnote{Very recently, a structured version of the RMR model, called \textsl{choice by salience}, has been proposed by \cite{GiarlottaPetraliaWatson2022b}.}
Rationalizable choice functions obviously belong to the first class of rationality, since a unique linear order suffices.  
On the other side of the scale of rationality, we find those choice functions that require the maximum number of rationales (namely $n-1$) to be justified. 
Despite its conceptually appealing motivation, the RMR model displays some drawbacks: (i) the family of rationalizing linear orders only provides a `non-structured' explanation of choice behavior; (ii) the class of maximally irrational choices (i.e., the ones requiring $n -1$ rationales) essentially collects all choices, even for very small sets of alternatives; and (iii) this model only applies to choice functions (but it could be naturally extended to choice correspondences).\footnote{The choice model based on salience \citep{GiarlottaPetraliaWatson2022b} creates a partition into $n$ classes of rationality, and positively addresses the first two issues of the RMR approach. Specifically, concerning (1), a binary relation of salience restricts the application of rationales to those indexed by the maximally salient items of a menu. Concerning (2), the smallest choice function in the last class of rationality that the authors are able to exhibit is defined on a set of 39 elements.}  

Another approach devoted to identify the degree of irrationality of a deterministic choice function is due to \cite{AmbrusRozen2014}.
As for the RMR model, also this approach is based on a counting technique.  
Specifically, the authors use a classical property of choice consistency ---namely \textsl{Independence of Irrelevant Alternatives} \citep{Arrow1950}, which is equivalent to \textsl{Axiom~$\alpha$} \citep{Chernoff1954} for choice functions--- to establish the degree of irrationality of a choice.
They count the number of violations of Axiom~$\alpha$ that a choice behavior exhibits: the larger this number, the less rational the behavior. 
In particular, they introduce a notion of violations of Axiom~$\alpha$, and accordingly define the \textsl{index of irrationality} of a choice by counting all menus that violate Axiom~$\alpha$. 
The abstract idea of their approach is appealing: it accounts to measure irrationality by counting deviations from rationality according to an axiomatic parameter (Axiom~$\alpha$).  

As we shall see, the approach developed in this paper measures the irrationality of choice behaviors in a way inspired by \cite{AmbrusRozen2014}.
In fact, similarly to them, we analyze deviations from rationality according to axiomatic parameters, namely Axioms~$\alpha$ and $\gamma$ \citep{Sen1971}, which are equivalent to the rationalizability of a quasi-choice. 
However, contrary to \cite{AmbrusRozen2014}, we do not \textit{directly} count violations of properties of choice consistency.
Instead, we use an \textit{indirect} approach: first we establish a theoretical way to measure violations, that is, a metric, and only then we count deviations from rationality using this metric. 
Of course, the soundness of such a procedure boils down to the selection of a metric that is both economically significant and highly discerning. 
The next three sections will extensively address this issue. 

%%%%%%%%%%%%%%%%%%%%%%%%%%%%%%%%
%%%%%%%%%%%%%%%%%%%%%%%%%%%%%%%%
%%%%%%%%%%%%%%%%%%%%%%%%%%%%%%%%
%%%%%%%%%%%%%%%%%%%%%%%%%%%%%%%%

\section{Metrics on quasi-choices} \label{SECT:metrics_on_choices}

This section is devoted to present ways to endow the family of all possible choice behaviors with metrics. 
%which are then used in Section~\ref{SECT:rational_index} to define a distance-based index of irrationality. 
%In studying the properties satisfied by these metrics, we shall identify those that are `undesirable' to measure deviations from  rationality. 
Specifically, first we recall a metric due to \cite{Klamler2008}, and then describe a variation of it, which showcases a rather sharp discernibility power.  
In Section~\ref{SECT:rational_index} we shall employ this novel metric as the measuring stick to evaluate deterministic deviations from rational behavior. 
%
%We clarify from the outset that the analysis developed here concerns the general case of quasi-choices. 
To start, we recall the notion of distance between quasi-choices.
 
\begin{definition} \rm \label{DEF:distance on choices}
    A \textsl{metric} on $\Choice(X)$ is a map $d \colon \Choice(X) \times \Choice(X) \to \R$ such that for all $C,C',C'' \in \Choice(X)$, the following properties hold:\vs
    \begin{itemize}
    \item[\textsf{[A0.1]}] $d(C,C') \geq 0$, and equality holds if and only if $C = C'$; 
    \item[\textsf{[A0.2]}] $d(C,C') = d(C',C)$; 
    \item[\textsf{[A0.3]}] $d(C,C') + d(C',C'') \geq d(C,C'')$.  
    \end{itemize}      
Property \textsf{A0.1} is \textsl{non-negativity}, property \textsf{A0.2} is \textsl{symmetry}, and property \textsf{A0.3} is the \textsl{triangle inequality}. 
\end{definition}

%Typically, a metric satisfies some additional properties, which make it well-suited for specific goals. 

%%%%%%%%%%%%%%%%%%%%%%%%%%%%%%%%%%%%%%
%%%%%%%%%%%%%%%%%%%%%%%%%%%%%%%%%%%%%%

\subsection{Klamler's metric}

The symmetric difference $\Delta$ of sets \citep{Kemeny1959} induces a metric on quasi-choices:  

\begin{definition}[\citealp{Klamler2008}] \rm \label{DEF:distance by symmetric difference}
    Let $d_\Delta \colon \Choice(X) \times \Choice(X) \to \R$ be the function defined as follows for all $C,C' \in \Choice(X)$:\vs
    $$
    d_\Delta(C,C') \coloneqq \sum_{S \in \XX} \big\vert C(S) \, \Delta \, C'(S) \big\vert \,.\vs
    $$    
\end{definition}

By Definition~\ref{DEF:distance by symmetric difference}, the distance between two quasi-choices over the same set of alternatives is obtained by a simple and intuitive procedure: first count the number of items in a menu that are in one choice set but not in the other one, and then take the sum of these numbers over all menus. 
Usually, being simple and intuitive is regarded as a good feature of a notion.  
Unfortunately, here this fact translates into an oversimplified evaluation of the distance between two behaviors, which totally neglects their structural features.  
Specifically, by only looking at the `size' of the disagreement of two quasi-choices over menus, Definition~\ref{DEF:distance by symmetric difference} fails to consider the `semantics' of this disagreement, which lies in the very nature of the items selected by exactly one of them.  
This in turn produces some important shortcomings of this metric in the process of detecting deviations from rational behavior.
The next two examples provide striking instances of this kind.  

\begin{example} \rm \label{EX:motivating}
	Consider the following three choice functions on $X=\{x,y,z\}$ (the unique item selected from each menu is underlined):\vs
	\begin{align*}
		(c_1) & \qquad \underline{x}y \,,\quad \underline{x}z \,,\quad \underline{y}z \,,\quad \underline{x}yz\,, \\
		(c_2) & \qquad \underline{x}y \,,\quad \underline{x}z \,,\quad \underline{y}z \,,\quad x\underline{y}z\,, \\
		(c_3) & \qquad \underline{x}y \,,\quad \underline{x}z \,,\quad \underline{y}z \,,\quad xy\underline{z}\,.
	\end{align*}
	The choices $c_1, c_2, c_3$ are equal on pairs of items but differ on the full menu $X$.
	On pairs, the selection process is reproduced by maximizing the linear order $x \succ y \succ z$. 
	However, $c_1$ is rationalizable by $\succ$, whereas $c_2$ and $c_3$ are not. 
	Intuition suggests that $c_3$ should be further than $c_2$ from the rational choice $c_1$. 
	For instance, if we use the linear order $\succ$ to rationalize pairs of items, then $c_2$ selects the second-best item from $X$, whereas $c_3$ ends up selecting the worst item of the three.\footnote{Of course, one may always consider different scenarios, in which $c_3$ is regarded more rational than $c_2$. However, these scenarios appear to be less likely to happen.}
	On the other hand, the metric $d_\Delta$ does regard $c_2$ and $c_3$ as equally distant from $c_1$, because we have\vs\vs
	$$
	d_\Delta(c_1,c_2) = \vert c_1(X) \Delta c_2(X) \vert = 2 \quad \text{ and } \quad d_\Delta(c_1,c_3) = \vert c_1(X) \Delta c_3(X) \vert = 2.
	$$
\end{example}   

\begin{example}\label{EX:metric_new} \rm
	Let $c_1,c_2,c_3,c_4$ be four choice correspondences over $X=\{x,y,z,w\}$, which are defined exactly in the same way for all menus distinct from $\{x,y,z\}$ and $X$, namely\vs\vs
	$$ 
	\underline{x}\underline{y} \,,\; \underline{x}z \,,\; \underline{x}w \,,\; \underline{y}\underline{z} \,,\; \underline{y}w \,,\; \underline{z}w \,,\; \underline{x}\underline{y}w \,,\; \underline{x}zw \,,\; \underline{y}\underline{z}w\,.\vs
	$$
	However, $c_1,c_2,c_3,c_4$ select different elements from the two menus $\{x,y,z\}$ and $X$, namely\vs
	\begin{itemize}
		\item[$(c_1)$] $\underline{x}\underline{y}z\,,\; \underline{x}\underline{y}zw\,$,\vs
		\item[$(c_2)$] $\underline{x}y\underline{z}\,,\; \underline{x}y\underline{z}w\,$,\vs
% 		\item[$(c_2^\prime)$] $xy\underline{z}\,,\; \underline{x}yzw\,$,\vs   
		\item[$(c_3)$] $x\underline{y}\underline{z}\,,\; x\underline{y}z\underline{w}\,$,\vs   
		\item[$(c_4)$] $x\underline{y}z\,,\; xyz\underline{w}\,$.\vs   
	\end{itemize} 
	Note that $c_1$ is rationalizable by the relation $\succ$ over $X$ such that $x \succ z$, $x \succ w$, $y \succ w$, and $z \succ w$.
	On the other hand, the three choices $c_2,c_3,c_4$ fail to be rationalizable, but they have exactly the same distance from the rationalizable choice $c_1$: 
	\begin{align*}
		d_\Delta(c_1,c_2) \;& =\;  \big\vert c_1(X) \,\Delta\, c_2(X) \big\vert + \big\vert c_1(\{x,y,z\}) \,\Delta\, c_2(\{x,y,z\}) \big\vert = 2 + 2 = 4 \,, \\
 		d_\Delta(c_1,c_3) \; & =\;  \big\vert c_1(X) \,\Delta\, c_3(X) \big\vert + \vert c_1(\{x,y,z\}) \,\Delta\, c_2(\{x,y,z\}) \vert = 2 + 2= 4\,,\\
		d_\Delta(c_1,c_4) \; & =\;  \big\vert c_1(X) \,\Delta\, c_3(X) \big\vert + \vert c_1(\{x,y,z\}) \,\Delta\, c_2(\{x,y,z\}) \vert = 3 + 1= 4\,.\vs  		
	\end{align*}
	However, similarly to Example~\ref{EX:motivating}, it is reasonable to assume that $c_2$ is `semantically' closer to $c_1$ than $c_3$ is: in fact, $c_2$ selects from the menus $\{x,y,z\}$ and $X$ some items that are better ranked (by $\succ$) than those selected by $c_3$. 
	There are also solid arguments to validate the opinion that $c_4$ should be the farthest choice from $c_1$. 
\end{example}

The low discernibility power of $d_\Delta$ is due to (some of) the properties it satisfies.
\cite{CarGiaWat2023} ---slightly correcting the findings of \cite{Klamler2008}--- prove that the following properties characterize $d_\Delta$ (a universal quantification is implicit): %: for any $C,C',C'',\widetilde{C}, \widetilde{C'} \in \Choice(X)$,
\begin{itemize}
    \item[\textsf{[A1]}] $d(C,C') + d(C',C'')  = d(C,C'')$ if and only if $C'$ is between $C$ and $C''$;\footnote{The notion of `betweenness' is due to \cite{AlabayrakAleskerov2000}: $C'$ is \textsl{between} $C$ and $C''$ if $C(S) \cap C''(S) \subseteq C'(S) \subseteq C(S) \cup C''(S)$ holds for any $S \in \XX$.}
  %  Then, $n \geq 3$ quasi-choices $C_1,C_2, \ldots, C_n \in \Choice(X)$ are \textsl{on a line} if, for all $i,j,k$ such that $1 \leq i < j < k \leq n$, $C_j$ is between $C_i$ and $C_k$. 
    \item[\textsf{[A2]}] if $\widetilde{C}$ and $\widetilde{C'}$ result from, respectively, $C$ and $C'$ by the same permutation of alternatives, then $d(C,C') = d(\widetilde{C},\widetilde{C'})$;
    \item[\textsf{[A3]}]  if $C$ and $C'$ agree on all (nonempty) menus in $\XX$ except for a subfamily $\XX' \subseteq \XX$, then the distance $d(C,C')$ is determined exclusively from the choice sets over $\XX'$;   
   % \item[\textsf{[A4]}] if $C,C',\widetilde{C},\widetilde{C'}$ only disagree on a menu $T \in \XX$ such that $C(T) = \widetilde{C}(T) \cup S$ and $C'(T) = \widetilde{C'}(T) \cup S$ for some $S \subseteq T$, then $d(C,C') = d(\widetilde{C},\widetilde{C'})$, 
    \item[\textsf{[A4$'$]}] if $C,C',\widetilde{C},\widetilde{C'}$ only disagree on a menu $T \in \XX$ such that $C(T) = \widetilde{C}(T) \Delta S$ and $C'(T) = \widetilde{C'}(T) \Delta S$ for some $S \subseteq T$, then $d(C,C') = d(\widetilde{C},\widetilde{C'})$;
    %\item[\textsf{[A5]}] $\min\{d(C,C'): C \neq C'\}=1$.
     \item[\textsf{[A5']}] for all $C \in \Choice(X)$ and $A \in \XX$, there is $C' \in \Choice(X)$ with the property that $\vert C(A) \Delta C'(A) \vert = 1 $, $C(B) = C'(B)$ for all $B \neq A$, and $d(C,C')=1$.
\end{itemize}

Axioms \textsf{A1}, \textsf{A2}, \textsf{A3}, \textsf{A4'}, and \textsf{A5'} are rather intuitive requirements for a metric on the family of quasi-choices.
In fact, Axiom~\textsf{A1} strengthens the triangle inequality \textsf{A0.3} by requiring that equality holds exactly for cases of betweenness. 
Axiom~\textsf{A2} states a condition of invariance under permutations. 
Axiom~\textsf{A3} is a separability property, whereas Axiom~\textsf{A4'} is a condition of translation-invariance.  
The first four properties produce a unique metric, up to some multiplicative coefficients that only depend on the size of the menu: Axiom~\textsf{A5'} forces these coefficients to be unique. 

As announced, we have: 

\begin{theorem}[\citealp{CarGiaWat2023}]\label{THM:main_theorem}
	The unique metric on $\Choice(X)$ that satisfies Axioms \textsf{A1},\,\textsf{A2},\,\textsf{A3},\,\textsf{A4'},\,\textsf{A5'} is $d_\Delta$. 
\end{theorem}

Unfortunately, the discernibility power of $d_\Delta$ among different choice behaviors is rather low, which is essentially due to the satisfaction of Axioms \textsf{A1} and \textsf{A3}.
To illustrate this fact, below we summarize some of the findings in \cite{CarGiaWat2023}.

\begin{definition} \rm \label{DEF:elementary_choices}
  A quasi-choice $C$ on $X$ is \textsl{elementary} if there is at most one menu $S \in \XX$ such that $C(S) \neq \varnothing$. 
  For any $S,T \in \XX \cup \{\es\}$ such that $T \subseteq S$, we denote by $C_{S \mapsto T}$ the elementary quasi-choice over $X$ defined as follows:\vs\vs
  $$
  C_{S \mapsto T}(A) \coloneqq 
  \left\{
  \begin{array}{lll}
  	T & \text{if } A =S \\
  	\varnothing & \text{otherwise.}
  \end{array}  
  \right.
  $$
\end{definition}

\begin{definition} \rm \label{DEF:characteristic_metric_on_sets}
Let $d$ be a metric on $\Choice(X)$, and $S \in \XX$.  
Denote by $\XX_S$ the family of all nonempty subsets of $S$.
Define a metric $d_S \colon \XX_S \cup \{\es\} \times \XX_S\cup \{\es\} \to \R$ by\vs
$$
d_S(A,B) \coloneqq d\left(C_{S \mapsto A},C_{S \mapsto B} \right)\vs
$$
for all $A,B \in \XX_S \cup \{\es\}$. %where $C_{S \mapsto A}$ is the elementary quasi-choice described in Definition~\ref{DEF:elementary_choices}.
We call $d_S$ the \textsl{characteristic metric induced by $d$ on $\XX_S \cup \{\es\}$}. 
\end{definition}

Any metric on $\Choice(X)$ satisfying \textsf{A1} and \textsf{A3} ---hence, in particular, $d_\Delta$--- is a sum of characteristic metrics:

\begin{lemma}[Elementary Decomposability] \label{LEMMA:semantics of A3}
Let $d$ be a metric on $\Choice(X)$ satisfying Axioms \textsf{A1} and \textsf{A3}.
For all $C,C' \in \Choice(X)$,\vs
$$
\displaystyle d(C,C') = \sum_{S \in \XX} d_S(C(S),C'(S)).
%d(C_{S \mapsto C(S)},C_{S \mapsto C'(S)}).
$$
\end{lemma}

The metric defined in the next section satisfies neither \textsf{A1} nor \textsf{A3}.

\subsection{A rational variation of Klamler's metric} \label{SECT:rational_metric}

We design a novel metric by suitably modifying Klamler's distance.   
This variation is inspired by \cite{AmbrusRozen2014}, because we employ two axioms of choice consistency ---in place of one--- to guide its construction:\vs 
\begin{description}
	\item[Axiom $\alpha\,$:]
	for all $A,B \subseteq X$ and $x \in X$, if $x \in A \subseteq B$ and $x \in C(B)$, then $x \in C(A)$;\vs
  	\item[Axiom $\gamma\,$:]
    for all $A,B \subseteq X$ and $x \in X$, if $x \in C(A)$ and $x \in C(B)$, then $x \in C(A \cup B)$.\vs
\end{description}

Axiom~$\alpha$ is due to \cite{Chernoff1954}.
In words, if an item is selected from a menu, then it is also chosen from any smaller menu containing it.
This property is often referred to as \textsl{Standard Contraction Consistency}. 
Its role in abstract theories of individual and social choice is central.  
\citet[p.\,407]{Nehring1997} even calls Axiom~$\alpha$ ``\textit{the mother of all choice consistency conditions'}'. 
Axiom~$\gamma$, often referred to as \textsl{Standard Expansion Consistency}, is due to \cite{Sen1971}.  
It says that if an item is selected from two menus, then it is also chosen from the larger menu obtained as their union. 

The connection between these two properties and rational behavior is well-known: 

\begin{theorem}[\citealp{Sen1971}]
	A choice correspondence is rationalizable if and only if both $\alpha$ and $\gamma$ hold.\footnote{This characterization readily extends to quasi-choices: see \citet[Theorem 2.5]{AizermanAleskerov1995}. For a proof of this generalization, see \citet[Theorem 2.8]{AleMon2002}.} 	
\end{theorem}

We now proceed to define a suitable refinement of $d_\Delta$, which takes into account all `locally rational approximations' of the original quasi-choice.
Specifically, we consider all restrictions of the given correspondence to all subsets of any given menu, and modify them in order to obtain quasi-choices that locally satisfy Axioms~$\alpha$ and~$\gamma$. 
Finally, we sum up all differences of these rational modifications. 

\begin{definition}\rm \label{DEF:alpha_closure_of_point}
	Let $C \colon \XX \cup \{\es\} \to \XX \cup \{\es\}$ be a quasi-choice over $X$, and $A \in \XX$ a nonempty menu. 
	Define a quasi-choice $C_A \colon \mathscr{A} \cup \{\es\} \to \mathscr{A} \cup \{\es\}$ over $A$, where $\mathscr{A}$ is the family of $\XX$ comprising all nonempty subsets of $A$, as follows for each $B \in \mathscr{A} \cup \{\es\}$:
	$$
	C_A(B) \coloneqq
	\left\{
	\begin{array}{ll}
		C(A)\cap B & \text{ if } C(A)\cap B \neq \varnothing, \\
		C(B) & \text{ otherwise.} 
	\end{array}\vs
	\right.
	$$
	We call $C_A$ the \textsl{rational localization of $C$ at $A$}.  
	Then, for all $C,C' \in \Choice(X)$, the \textsl{rational distance} between $C$ and $C'$ is defined by\vs 
	$$
	d_\rat(C,C')\coloneqq \sum_{A \in \XX}  d^A_\Delta(C_A,C'_A),\vs\vs
	$$
	where $d^A_\Delta$ denotes the restriction $d_\Delta \!\!\upharpoonright_{\Choice(A) \times \Choice(A)}$.  
\end{definition}

Definition~\ref{DEF:alpha_closure_of_point} employs $(\vert 2^X \vert -1)$-many restrictions of the given metric $d_\Delta$ to  compare standard modifications of two given quasi-choices: these modifications are a sort of rational closures of a given choice on a given menu.
Note that the terminology of `local rationalization' used for $C_A$ is motivated by the fact that any element $x \in C(A)$ is never responsible for a violation of Axioms~$\alpha$ or~$\gamma$ by $C_A$.\footnote{More formally, considering Axiom~$\alpha$, this means that if $y \in S \subseteq T \subseteq A$, $y \in C_A(T)$, and $y \notin C_A(S)$, then $y \notin C(A)$. Similarly, for Axiom~$\gamma$, if $S,T \subseteq A$, $y \in C_A(S) \cap C_A(T)$, and $y \notin C_A(S \cup T)$, then $y \notin C(A)$.}  

\begin{example} \label{EX:computation C_A} \rm 
	We illustrate how Definition~\ref{DEF:alpha_closure_of_point} works in a very simple case.
	Consider the two choice functions $c_2$ and $c_2^\prime$ over $X= \{x,y,z\}$ defined by\vs\footnote{The choice function $c_2$ has already been considered in Example~\ref{EX:motivating}.}
	\begin{align*}
		(c_2) & \qquad \underline{x}y \,,\quad \underline{x}z \,,\quad \underline{y}z \,,\quad x\underline{y}z\,, \\
		(c_2^\prime) & \qquad x\underline{y} \,,\quad \underline{x}z \,,\quad \underline{y}z \,,\quad x\underline{y}z\,.
	\end{align*}
	Note that $c_2$ and $c_2^\prime$ are equal except on the menu $\{x,y\}$, and $c_2^\prime$ is rationalizable by the linear order $y \succ x \succ z$. 
	To determine $d_\rat(c_2,c_2^\prime)$, we preliminary compute their rational localizations $(c_2)_A$ and $(c_2^\prime)_A$ at any nonempty subset $A$ of $X$ having size at least two: 
	\begin{table}[h!]
	\begin{center}	\small
	\begin{tabular}{|l||*{2}{c|}}
	\hline
	\backslashbox{menu}{localization}
	&\makebox[2em]{$(c_2)_A \in \choice(A)$}&\makebox[2em]{$(c_2^\prime)_A \in \choice(A)$} \\
	\hline\hline  
	\small $A =\{x,y\}$ & $\underline{x}\,,\;\underline{y}\,,\;\underline{x}y$ & $\underline{x}\,,\;\underline{y}\,,\;x\underline{y}$ \\ 
	\hline
	\small $A =\{x,z\}$ & $\underline{x}\,,\;\underline{z}\,,\;\underline{x}z$ & $\underline{x}\,,\;\underline{z}\,,\;\underline{x}z$ \\ 
	\hline
	\small $A =\{y,z\}$ & $\underline{y}\,,\;\underline{z}\,,\;\underline{y}z$ & $\underline{y}\,,\;\underline{z}\,,\;\underline{y}z$ \\ 
	\hline
	\small $A =\{x,y,z\}$ & $\underline{x}\,,\;\underline{y}\,,\;\underline{z}\,,\;x\underline{y}\,,\;\underline{x}z\,,\;\underline{y}z\,,\;x\underline{y}z$ & $\underline{x}\,,\;\underline{y}\,,\;\underline{z}\,,\;x\underline{y}\,,\;\underline{x}z\,,\;\underline{y}z\,,\;x\underline{y}z$ \\ 
	\hline
	\end{tabular}
	\end{center}
	\vs\vs\vs\vs\vs
	\end{table}

\noindent 
(Note that all rational localizations at singletons are trivial choice functions in this particular case.) 
Since $(c_2)_A=(c^\prime_2)_A$ for each $A\neq \{x,y\}$, whereas $(c_2)_{\{x,y\}}(\{x,y\})=x$ and $(c_2^\prime)_{\{x,y\}}(\{x,y\})=y$, we conclude\vs\vs 
$$
d_\rat(c_2,c_{2}^\prime)=d^{\{x,y\}}_{\Delta}\big((c_2)_{\{x,y\}},(c_2^\prime)_{\{x,y\}} \big)=2\,.
$$
\end{example}

\smallskip

As possibly expected, Definition~\ref{DEF:alpha_closure_of_point} is sound: 

\begin{lemma} \label{LEMMA:new_metric_is_sound}
	The function $d_\rat$ is a metric on $\Choice(X)$.
\end{lemma}

\begin{proof} 
	For \textsf{A0.1}, clearly $d_\rat(C,C')$ is nonnegative. 
	If $d(C,C')=0$, then $d^A_\Delta(C_A,C'_A)=0$ for all $A \in \XX$.
	It follows that $C_A(B)=C'_A(B)$ for all $B \subseteq A$, and so $C(A)=C_A(A)=C'_A(A)=C'(A)$.
	This proves \textsf{A0.1}.
	Axiom \textsf{A0.2} is obvious. 
	For \textsf{A0.3}, observe that $d^A_\Delta(C_A,C''_A) \leq d^A_\Delta(C_A,C'_A) + d^A_\Delta(C'_A,C''_A)$, because $d^A_\Delta$ is the restriction of a metric. 
	Thus the claim follows from summing over all $A \in \XX$.
    \qed
\end{proof}

The next remark shows that, despite being derived from $d_\Delta$, the rational metric $d_\rat$ does not satisfy several properties considered by Klamler; in particular, neither of the two axioms responsible for elementary decomposability ---namely \textsf{A1} and \textsf{A3}--- hold for $d_\rat$.

\begin{remark} \rm \label{REM:nontriviality_new_metric}
We prove that $d_\rat$ satisfies neither \textsf{A1} nor \textsf{A3} nor \textsf{A4'}.
All counterexamples will be quasi-choices over the set $X= \{x,y,z\}$. 
Since in all cases the choice set of any singleton is nonempty, we only define them on menus having size two or three.
\medskip
  
To prove the failure of \textsf{A1}, define $C,C',C'' \in \Choice(X)$ by\vs\vs
\begin{itemize}
	\item[\rm $(C)$] $\underline{x}y \,,\; \underline{x}\underline{z} \,,\; yz \,,\; \underline{x}y\underline{z}\,$;\vs
	\item[\rm $(C')$] $\underline{x}\underline{y} \,,\; \underline{x}\underline{z} \,,\; \underline{y}z \,,\; \underline{x}\underline{y}z\,$;\vs
	\item[\rm $(C'')$] $\underline{x}\underline{y} \,,\; \underline{x}\underline{z} \,,\; \underline{y}\underline{z} \,,\; x\underline{y}z\,$.\vs
\end{itemize}
Clearly, $C'$ is between $C$ and $C''$.
However, $d_\rat(C,C'')=10$ is different from $d_\rat(C,C') + d_\rat(C',C'') = 8+4 =12$.
\medskip

For the failure of \textsf{A3}, define $C,C',D,D' \in \Choice(X)$ by\vs\vs
\begin{itemize}
	\item[\rm $(C)$] $\underline{x}y \,,\; xz \,,\; y\underline{z} \,,\; \underline{x}\underline{y}z\,$;\vs
	\item[\rm $(C')$] $xy \,,\; xz \,,\; y\underline{z} \,,\; \underline{x}\underline{y}z\,$;\vs
	\item[\rm $(D)$] $\underline{x}y \,,\; xz \,,\; yz \,,\; xy\underline{z}\,$;\vs
	\item[\rm $(D')$] $xy \,,\; xz \,,\; yz \,,\; xy\underline{z}\,$.\vs
\end{itemize}	
 Let $\XX'=\{\{x,y\},\{x,z\}\}$, and observe that $C \!\!\upharpoonright_{\XX \setminus \XX'}=C'\!\!\upharpoonright_{\XX \setminus \XX'}$, $D \!\!\upharpoonright_{\XX \setminus \XX'}=D'\!\!\upharpoonright_{\XX \setminus \XX'}$, $C \!\!\upharpoonright_{ \XX'}=D \!\!\upharpoonright_{ \XX'}$, and $C'\!\!\upharpoonright_{ \XX'}=D'\!\!\upharpoonright_{ \XX'}$.
 However, $d_\rat(C,C')=1$ whereas $d_\rat(D,D')=2$.
 \medskip

 For the failure of \textsf{A4'}, define $C,\widetilde{C},C',\widetilde{C'} \in \Choice(X)$ by\vs\vs
\begin{itemize}
	\item[\rm $(C)$] $\underline{x}y \,,\; \underline{x}z \,,\; \underline{y}z \,,\;\underline{x}yz\,$;\vs
	\item[\rm $(\widetilde{C})$] $\underline{x}y \,,\; \underline{x}z \,,\; \underline{y}z \,,\; \underline{x}\underline{y}\underline{z}\,$;\vs
	\item[\rm $(C')$] $\underline{x}y \,,\; \underline{x}z \,,\; \underline{y}z \,,\; x\underline{y}\underline{z}\,$;\vs
	\item[\rm $(\widetilde{C'})$] $\underline{x}y \,,\; \underline{x}z \,,\; \underline{y}z \,,\; xyz\,$.\vs
\end{itemize}
The four quasi-choices over $X$ agree on every menu, except on $X$.
For $S=\{y,z\}$, we have $C(X)=\widetilde{C}(X)\Delta S$ and $C'(X)=\widetilde{C'}(X)\Delta S$, and yet $d_\rat(C,C')=8 \neq 6 = d_\rat(\widetilde{C},\widetilde{C'})$.
\end{remark}

It would be interesting to axiomatically characterize the rational metric $d_\rat$: we leave this as an open problem. 
%\textbf{\magenta (Maybe we should prove a characterization, if we wish to submit it to JET)}

%%%%%%%%%%%%%%%%%%%%%%%%%%%%%%%%%%%
%%%%%%%%%%%%%%%%%%%%%%%%%%%%%%%%%%%
%%%%%%%%%%%%%%%%%%%%%%%%%%%%%%%%%%%
%%%%%%%%%%%%%%%%%%%%%%%%%%%%%%%%%%%

\section{Distance-based degrees of irrationality} \label{SECT:rational_index}

We finally give a formal definition of the measure of irrationality of a deterministic choice behavior with respect to a given metric, where the family of rationalizable quasi-choices acts as the benchmark of rationality. 
We provide two versions of it: (1) simple, and (2) weighted. 
The first applies to all quasi-.choices, whereas the second is only designed for choice correspondences. 

%%%%%%%%%%%%%%%%%%%%%%%%%%%%%%%%%%%%%%%
%%%%%%%%%%%%%%%%%%%%%%%%%%%%%%%%%%%%%%%

\subsection{A simple degree of irrationality}

\begin{definition} \rm \label{DEF:irrationality_degree}
	Let $\rho \colon \Choice(X) \to \Choice(X)$ be a metric. 
	Denote by $\Choice_{\mathrm{rat}}(X)$ the subfamily of $\Choice(X)$ comprising all quasi-choices that are rationalizable.   
	For any quasi-choice $C$ over $X$, the $\rho$\textsl{-degree of irrationality} of $C$ is the integer defined by\vs
	$$
	\irr_\rho (C) \coloneqq \min \{\rho(C,D) : D \in \Choice_{\mathrm{rat}}(X)\}.\vs 
	$$ 
	(This degree is well-defined, because $X$ is finite.)
\end{definition}

Given a metric $\rho$ on $\Choice(X)$, the larger the $\rho$-degree of irrationality of a quasi-choice $C$ is, the more irrational $C$ is considered \textit{from the point of view of $\rho$}. 
Note that if a quasi-choice is rationalizable, then its $\rho$-degree of irrationality is zero for any metric $\rho$.   
For instance, the choice function $c_1$ defined in Example~\ref{EX:motivating} has a $d_\Delta$-degree of irrationality equal to zero, whereas $c_2$ and $c_3$ have a $d_\Delta$-degree of irrationality equal to two.  

As already pointed out, the soundness of Definition~\ref{DEF:irrationality_degree} depends on the economic significance and the discernibility power of the metric used to determine the degree of irrationality. 
In this respect, the rational metric $d_\rat$ appears to be better suited than Klamler's distance $d_\Delta$.
The next two examples witness this claim. 

\begin{example} \rm \label{EX:3_choice_continued}
	Consider the three choice functions $c_1,c_2,c_3$ defined in Example~\ref{EX:motivating}.
	It is easy to show that\vs 
	$$
	\irr_{d_\Delta}(c_1) = 0\,,\quad \irr_{d_\Delta}(c_2)= 2\,,\quad \irr_{d_\Delta}(c_3)=2\,.\vs 
	$$
	On the other hand, below we show that\vs 
	$$
	\irr_{d_\rat}(c_1) = 0\,,\quad \irr_{d_\rat}(c_2)= 2\,,\quad \irr_{d_\rat}(c_3)=3\,.\vs 
	$$
	\begin{description}
	 \item[$\bullet$ $\irr_{d_\rat}(c_1)= 0$:] This is obvious, because $c_1$ is rationalizable.
	 
	 \item[$\bullet$ $\irr_{d_\rat}(c_2)= 2$:] As noted in Example~\ref{EX:computation C_A}, the choice function $c_2^\prime$ defined by\vs 
	$$
	\qquad x\underline{y} \,,\quad \underline{x}z \,,\quad \underline{y}z \,,\quad x\underline{y}z\vs
	$$
	is rationalizable by the linear order $\succ$, with $y \succ x \succ z$.
	We know that $d_\rat(c_2,c_{2}^\prime)=2$.
	Therefore, to prove the claim, we show that $d_\rat(c_2,D) \geq 2$ for all $D \in \Choice_{\mathrm{rat}}(X)$.
	
	Hereafter we shall employ a simplified notation, which is also used -- \textit{mutatis mutandis} -- in the proof of the equality $\irr_{d_\rat} (c_3) =3$.
	Specifically, for all $A \in \XX$, we denote $d^{A}_\Delta((c_2)_{A},D_A)$ by the less cumbersome $d^{A}_\Delta$. 
	Moreover, we drop brackets and set separators whenever clear from context, using $D(xz)=x$ instead of $D(\{x,z\})=\{x\}$, $d^{xz}_\Delta$ instead of $d^{\{x,z\}}_\Delta$, etc.
	
	Now fix $D \in \Choice_{\mathrm{rat}}(X)$.
	Then, either (1) $y \in D(xy)$, or (2) $y \notin D(xy)$. \vs
	\begin{description}
		\item [\rm(1)] If $y \in D(xy)$, then we separately consider two cases.
		\begin{description}
		    \item [\rm(1A)] If $x \notin D(xy)$, then $D_{xy}(xy)=y$.
		    Since $(c_2)_{xy}(xy)=x$, we obtain $d_{\Delta}^{xy}\geq 2$, hence $d_\rat(c_2,D)\geq 2$.
		    \item [\rm(1B)] If $x \in D(xy)$, then we split the analysis in two subcases.
		    \begin{description}
		        \item [\rm(1B1)] If $x \notin D(xz)$, then $x \notin D_{xz}(xz)$, while $x \in (c_2)_{xz}(xz)$.
		        It follows that $d_{\Delta}^{xz} \geq 1$.
		        Note that $y \in D_{xy}(xy)$ and $y \notin (c_2)_{xy}(xy)$ imply $d_{\Delta}^{xy} \geq 1$.
		        We conclude $d_\rat(c_2,D) \geq 2$.
		        \item [\rm(1B2)] If $x \in D(xz)$, then $x \in D(xyz)$ by Axiom~$\gamma$, and so $x \in D_{xyz}(xyz)$.
		        Since $x \notin (c_2)_{xyz}(xyz)$, we get $d_{\Delta}^{xyz} \geq 1$.
		        As before, $y \in D_{xy}(xy)$ and $y \notin (c_2)_{xy}(xy)$ imply $d_{\Delta}^{xy} \geq 1$.
		        We conclude $d_\rat(c_2,D) \geq 2$.
		    \end{description}
		\end{description}
		\item [\rm(2)] If $y \notin D(xy)$, then $y \notin D(xyz)$ by Axiom~$\alpha$.
		Thus $y \notin D_{xyz}(xyz) \cup D_{xyz}(xy)$ and $y \in (c_2)_{xyz}(xyz) \cap (c_2)_{xyz}(xy)$.
		We conclude $d_{\Delta}^{xyz} \geq 2$, hence $d_\rat(c_2,D) \geq 2$.
	\end{description}
	
	\item[$\bullet$ $\irr_{d_\rat}(c_3)= 3$:]
	Let $c_3^\prime$ be the choice rationalizable by the relation $\succ$ on $X$ defined by $z\succ x$ and $x \succ y$, that is,\vs 
	$$ 
	\underline{x}y \,,\; x\underline{z} \,,\; \underline{y}\underline{z} \,, \, xy\underline{z}\,.
	$$ 
	(Note that $\succ$ is not transitive, because $y$ and $z$ are incomparable.) 
	It is easy to check that $d_\Delta^{xz}((c_3)_{xz},(c_3^\prime)_{xz})=2$ and $d_\Delta^{yz}((c_3)_{yz},(c_3^\prime)_{yz})=1$, whereas all other rational localizations of $c_3$ and $c_3^\prime$ coincide. 
	It follows that $d_\rat(c_3,c_3')=3$.
	
	To complete the proof, we show that $d_\rat(c_3,D) \geq 3$ for any $D \in \Choice_{\mathrm{rat}}(X)$.\vs
	\begin{description}
		\item [\rm (1)] If $z \notin D(xyz)$, then either (1A) $z \notin D(xz)$, or (1B) $z \notin D(yz)$, using Axiom~$\gamma$.
	    \begin{description}
	        \item [\rm(1A)] If $z \notin  D(xz)$, then we split the analysis into two subcases.
	        \begin{description}
	            \item [\rm(1A1)] If $x  \in D(xyz)$, then $x \in D_{xyz}(xyz)$, $z \notin D_{xyz}(xyz)$, and $D_{xyz}(xz) = x$.
	            Therefore, from $z \in (c_3)_{xyz}(xyz) \cap (c_3)_{xyz}(xz)$ and  $x \notin (c_3)_{xyz}(xz)$, we derive  $d_\Delta^{xyz} \geq 3$. We conclude $d_\rat(c_3,D) \geq 3$.
	            \item [\rm(1A2)] If $x \notin D(xyz)$, then by Axiom~$\gamma$, $x \notin D(xy)$ or $x \notin D(xz)$, and so $x \notin D_{xy}(xy)$ or $x \notin D_{xz}(xz)$.
	            Since $x \in (c_3)_{xy}(xy) \cap (c_3)_{xz}(xz)$, we obtain $d_{\Delta}^{xy} \geq 1$ or $d_{\Delta}^{xz} \geq 1$.
	            Finally, since $z \in (c_3)_{xyz}(xyz) \cap (c_3)_{xyz}(xz)$ and $z \notin D_{xyz}(xyz) \cup D_{xyz}(xz)$, we get $d_{\Delta}^{xyz}\geq 2$, and so $d_\rat(c_3,D) \geq 3$.
	        \end{description}
	        \item [\rm(1B)] Suppose $z \notin  D(yz)$. We can assume that $z \in D(xz)$, otherwise we would be done by case (1A).
	        It follows that $z \in D_{xz}(xz)$ while $z \notin c_{3_{xz}}(xz)$, hence $d_{\Delta}^{xz}\geq 1$.
	        Since $z \notin D_{xyz}(xyz)$ and $z \notin D_{xyz}(yz)$, we conclude $d_{\Delta}^{xyz} \geq 2$, and so $d_{\rat}(c_3,D) \geq 3$.
	    \end{description}
	    \item [\rm(2)] If $z \in D(xyz)$, then $z \in D(xz)$ and $z \in D(yz)$ by Axiom~$\alpha$.
	    \begin{description}
	        \item [\rm(2A)] If $x \notin D(xy)$, then $d_\Delta^{xy}\geq 1$, $d_\Delta^{xz}\geq 1$, $d_\Delta^{yz}\geq 1$, and so $d_\rat(c_3,D) \geq 3$.
	        \item [\rm(2B)] If $x \in D(xy)$, we consider two subcases.
	        \begin{description}
	            \item [\rm(2B1)] If $x \notin D(xyz)$, then $x \notin D(xz)$ by Axiom~$\gamma$, hence  $d_\Delta^{xz}\geq 2$ and $d_\Delta^{yz}\geq 1$. We conclude $d_\rat(c_3,D) \geq 3$.
	            \item [\rm(2B2)] If $x \in D(xyz)$, then $x \in D_{xyz}(xz)$ and $x \in D(xz)$ by Axiom~$\alpha$, hence $d_\Delta^{xyz} \geq 2$ and $d_\Delta^{xz}=1$.
	                Again, we conclude $d_\rat(c_3,D) \geq 3$.
	        \end{description}
	    \end{description}
	\end{description}
	\end{description}
\end{example}

\begin{example}\label{EX:metric_new_continued} \rm
	Let $c_1,c_2,c_3,c_4$ be the four choice correspondences over $X=\{x,y,z,w\}$ defined in Example~\ref{EX:metric_new}. 
	One can easily show that these three choices have the same $d_\Delta$-degree of irrationality, being\vs 
	\begin{equation*}
		\irr_{d_\Delta}(c_2) = 
		\irr_{d_\Delta}(c_3) = 
		\irr_{d_\Delta}(c_4) = 4\,.\vs  		
	\end{equation*}
	On the contrary, the metric $d_\rat$ agrees with the perception that $c_2$ is less irrational than $c_3$, and $c_4$ is the most irrational of all, being\vs 
	\begin{equation} \label{EQ:irrationality_of_choices_on_4_elements}
		\irr_{d_\rat} (c_1) = 0\,,\;\; \irr_{d_\rat} (c_2) = 6\,,\;\; \irr_{d_\rat}(c_3) = 16\,,\;\; \irr_{d_\rat} (c_4) = 19\,.\vs
	% \irr_{d_\rat} (c_2^\prime) = 9		
	\end{equation}
	The related computations are extremely long and tedious, so we omit them.\footnote{However, they are available upon request.} 
\end{example}

%%%%%%%%%%%%%%%%%%%%%%%%%%%%%%%%%%%%%%%
%%%%%%%%%%%%%%%%%%%%%%%%%%%%%%%%%%%%%%%

\subsection{A weighted degree of irrationality}

The evaluation of the degree of irrationality of choice behavior described above can be refined, as long as the DM is able to provide additional pieces of information. 
Here we illustrate a possible refinement of it, which applies to the family of choice correspondences; in other words, we consider the special case of a decisive DM.

In a preliminary step, the DM is required to provide additional information about the `subjective desirability' of all rational choice behaviors.
Operationally, this is obtained by assigning weights to each rationalizable choice correspondence. 
According to intuition, very desirable rational behaviors should be given a weight less or equal than one, because this may produce the effect of contracting the rational distance of all choices close to them.
On the contrary, less appealing rational behaviors should given a weight greater or equal than one, in order to possibly dilate the distance from rationalization.  
Once desirability is assessed, the irrational degree of a decisive choice behavior is then computed as the minimum weighted distance from the benchmark of rationality.  

In the process of designing the weighting procedure, we adhere to some natural rules of conduct. 
We select the transitivity of the relation of revealed preference as our guiding parameter: the more transitive this relation is, the more desirable the associated behavior becomes, and the lower the corresponding weight must be. 
From this point of view, the most desirable choices will be those rationalized by \textsl{weak orders} (asymmetric and negatively transitive, hence transitive), which will be assigned the lowest weight among all rational behaviors.   
Less desirable levels are those of choices rationalized by \textsl{semiorders} (asymmetric, Ferrers, and semitransitive) \citep{Luce1956}, and by \textsl{interval orders} (asymmetric and Ferrers)  \citep{Fishburn1970,Fishburn1985}.  
At an even lower desirability level lie all choices rationalized by transitive asymmetric relations that fail to be interval orders. 
At the bottom of the scale, we find those choices that are rationalized by asymmetric, acyclic and intransitive binary relations, which will be given the highest weight of all. 

An even finer tuning of the weighting procedure can be achieved by employing the so-called \textsl{strict and weak $(m,n)$-Ferrers properties} \citep{GiaWat2014Ferrers,GiaWat2018}, which provide a classification of all asymmetric and acyclic binary relations on a set according to their discrete level of transitivity.\footnote{On the point, see also \cite{CanGiaGreWat2016} for a classification of all rationalizable choices on the basis of the so-called \textsl{axioms of $(m,n)$-replacement consistency}.} 
The next definition provides a simplified version of these properties, which is however sufficient for our goal. 

\begin{definition}[\citealp{GiaWat2014Ferrers}] \rm \label{DEF:Ferrers_properties}
Let $\succ$ be an asymmetric and acyclic binary relation over $X$.
Denote by $\succsim$ the \textsl{canonical completion} of $\succ$, obtained by adding all $\succ$-incomparable pairs to $\succ$.\footnote{Two (not necessarily distinct) elements $x,y \in X$ are \textsl{$\succ$-incomparable} if nether $x \succ y$ nor $y \succ x$ holds. Technically, the canonical completion $\succsim$ is the extension of $\succ$ in which incomparability is transformed into indifference. In particular, the canonical completion $\succsim$ of $\succ$ is both \textsl{reflexive} (i.e., $x \succsim x$ for all $x \in X$) and \textsl{complete} (i.e., $x \succsim y$ or $y \succsim x$ for all distinct $x,y \in X$).}  
For any integers $m \geq n \geq 1$, we say that $\succ$ is \textit{$(m,n)$-Ferrers} if the joint satisfaction of $(x_1 \succsim \ldots \succsim x_m)$ and $(y_1 \succsim \ldots \succsim y_n)$ implies either $x_1 \succsim y_n$ or $y_1 \succsim x_m$, for all (not necessarily distinct) $x_1, \ldots, x_m, y_1, \ldots, y_n \in X$. 
\end{definition}

It is easy to show that $(m,n)$-Ferrers implies $(m',n')$-Ferrers for any $1 \leqslant m' \leqslant m$ and $1 \leqslant n' \leqslant n$ \citep[Lemma~2.6]{GiaWat2014Ferrers}.
Furthermore, $(3,3)$-Ferrers implies $(m,n)$-Ferrers for any $m\geqslant n \geqslant 1$ \citep[Theorem~3.1(v)]{GiaWat2014Ferrers}.
Note also that $(3,3)$-Ferrers relations are weak orders, $(3,1)$- and $(2,2)$-Ferrers relations are semiorders, $(2,2)$-Ferrers relations are interval orders, $(2,1)$-Ferrers relations are transitive, and $(1,1)$-Ferrers are acyclic but intransitive.  

\begin{figure}[h]
\begin{center}
\psset{xunit=2.2cm} \psset{yunit=2.3cm}
\begin{pspicture}[showgrid=false](0,0.5)(4,10)
          \psset{linewidth=0.5pt}
%%%%%%   labels          
\rput(-1,9.75){\textbf{Desirability class}}     
\rput(2,9.75){\textbf{Asymmetric and acyclic relations}}
\rput(-1,9){\large \textbf{(1)}}     
\rput(-1,8){\large \textbf{(2)}}     
\rput(-1,7){\large \textbf{(3)}}     
\rput(-1,6){\large \textbf{(4)}}     
\rput(-1,5){\large \textbf{(5)}}     
\rput(-1,4){\large \textbf{(6)}}     
\rput(-1,3){\large \textbf{(7)}}     
\rput(-1,2){\large \textbf{(8)}}
\rput(-1,1){\large \textbf{(9)}}          
% lines
% (3,3) to (4,2)
          \psline[linecolor=darkgray,linewidth=0.07,arrowsize=0.3]{->}(2,8.8)(2,8.21)
% (4,2) to (5,1)-(3,2)
          \psline[linecolor=darkgray,linewidth=0.07,arrowsize=0.3]{->}(2,7.8)(2,7.24)
% (5,1)-(3,2) to (5,1)-(2,2)
          \psline[linecolor=darkgray,linewidth=0.07,arrowsize=0.3]{->}(1.8,6.75)(1.15,6.23)
% (5,1)-(3,2) to (4,1)-(3,2)
          \psline[linecolor=darkgray,linewidth=0.07,arrowsize=0.3]{->}(2.2,6.75)(2.85,6.23)
% (5,1)-(2,2) to (5,1)
          \psline[linecolor=darkgray,linewidth=0.07,arrowsize=0.3]{->}(0.68,5.75)(0.02,5.21)
% (4,1)-(3,2) to (3,2)
          \psline[linecolor=darkgray,linewidth=0.07,arrowsize=0.3]{->}(3.3,5.75)(3.96,5.21)  
% (5,1)-(2,2) to (4,1)-(2,2)
          \psline[linecolor=darkgray,linewidth=0.07,arrowsize=0.3]{->}(1.57,5.78)(1.89,5.21)
% (4,1)-(3,2) to (4,1)-(2,2)
          \psline[linecolor=darkgray,linewidth=0.07,arrowsize=0.3]{->}(2.43,5.78)(2.12,5.21)
% (4,1)-(2,2) to (4,1)
          \psline[linecolor=darkgray,linewidth=0.07,arrowsize=0.3]{->}(1.57,4.8)(0.28,4.19)
% (4,1)-(2,2) to (3,1)-(2,2)
          \psline[linecolor=darkgray,linewidth=0.07,arrowsize=0.3]{->}(2.43,4.8)(3.44,4.22)          
%
%  eliminati
%          
% (5,1)-(2,2) to (3,1)-(2,2)
%          \psline[linecolor=darkgray,linewidth=0.07,arrowsize=0.3]{->}(1.57,5.78)(3.43,4.22)
% (4,1)-(3,2) to (4,1)
%          \psline[linecolor=darkgray,linewidth=0.07,arrowsize=0.3]{->}(2.43,5.78)(0.28,4.18)
%
%  fine eliminati
%
% (5,1) to (4,1)
          \psline[linecolor=darkgray,linewidth=0.07,arrowsize=0.3]{->}(0,4.8)(0,4.21)
% (3,2) to (3,1)-(2,2)
          \psline[linecolor=darkgray,linewidth=0.07,arrowsize=0.3]{->}(4,4.8)(4,4.24)
% (4,1) to (3,1)
          \psline[linecolor=darkgray,linewidth=0.07,arrowsize=0.3]{->}(0.08,3.79)(0.85,3.2)
% (3,1)-(2,2) to (2,2)
          \psline[linecolor=darkgray,linewidth=0.07,arrowsize=0.3]{->}(3.92,3.77)(3.15,3.2)
% (3,1)-(2,2) to (3,1)
          \psline[linecolor=darkgray,linewidth=0.07,arrowsize=0.3]{->}(3.42,3.79)(1.27,3.18)
% (3,1) to (2,1)
          \psline[linecolor=darkgray,linewidth=0.07,arrowsize=0.3]{->}(1.13,2.79)(1.87,2.2)
% (2,2) to (2,1)
          \psline[linecolor=darkgray,linewidth=0.07,arrowsize=0.3]{->}(2.87,2.79)(2.13,2.2)
% (2,1) to (1,1)
          \psline[linecolor=darkgray,linewidth=0.07,arrowsize=0.3]{->}(2,1.8)(2,1.21)
%
% Ferrers properties
          \psline[fillstyle=solid,fillcolor=lightgray](1.75,8.83)(2.25,8.83)(2.25,9.17)
          (1.75,9.17)(1.75,8.83)
          \rput(2,9){$(3,3)$}
          \rput(2.73,9){\small \emph{weak order}}
          \psline[fillstyle=solid,fillcolor=lightgray](1.75,7.83)(2.25,7.83)(2.25,8.17)
          (1.75,8.17)(1.75,7.83)
          \rput(2,8){$(4,2)$}
          \psline[fillstyle=solid,fillcolor=lightgray](1.45,6.8)(2.55,6.8)(2.55,7.2)
          (1.45,7.2)(1.45,6.8)
          \rput(2,7){$(5,1) \:\&\: (3,2)$}
          \psline[fillstyle=solid,fillcolor=lightgray](0.45,5.8)(1.55,5.8)(1.55,6.2)
          (0.45,6.2)(0.45,5.8)
          \rput(1,6){$(5,1) \:\&\: (2,2)$}
          \psline[fillstyle=solid,fillcolor=lightgray](2.45,5.8)(3.55,5.8)(3.55,6.2)
          (2.45,6.2)(2.45,5.8)
          \rput(3,6){$(4,1) \:\&\: (3,2)$}
          \rput(4.27,6){\small \emph{strong semiorder}}
          \psline[fillstyle=solid,fillcolor=lightgray](-0.25,4.83)(0.25,4.83)(0.25,5.17)
          (-0.25,5.17)(-0.25,4.83)
          \rput(0,5){$(5,1)$}
          \psline[fillstyle=solid,fillcolor=lightgray](3.75,4.83)(4.25,4.83)(4.25,5.17)
          (3.75,5.17)(3.75,4.83)
          \rput(4,5){$(3,2)$}
          \rput(5.1,5){\small \emph{strong interval order}}
          \psline[fillstyle=solid,fillcolor=lightgray](1.45,4.83)(2.55,4.83)(2.55,5.17)
          (1.45,5.17)(1.45,4.83)
          \rput(2,5){$(4,1) \:\&\: (2,2)$}
          \psline[fillstyle=solid,fillcolor=lightgray](-0.25,3.83)(0.25,3.83)(0.25,4.17)
          (-0.25,4.17)(-0.25,3.83)                    
          \rput(0,4){$(4,1)$}
          \psline[fillstyle=solid,fillcolor=lightgray](3.45,3.8)(4.55,3.8)(4.55,4.2)
          (3.45,4.2)(3.45,3.8)
          \rput(4,4){$(3,1) \:\&\: (2,2)$}
          \rput(5,4){\small \emph{semiorder}}
          \psline[fillstyle=solid,fillcolor=lightgray](0.75,2.83)(1.25,2.83)(1.25,3.17)
          (0.75,3.16)(0.75,2.83)
          \rput(1,3){$(3,1)$}
          \psline[fillstyle=solid,fillcolor=lightgray](2.75,2.83)(3.25,2.83)(3.25,3.17)
          (2.75,3.17)(2.75,2.83)
          \rput(3,3){$(2,2)$}
          \rput(3.84,3){\small \emph{interval order}}
          \psline[fillstyle=solid,fillcolor=lightgray](1.75,1.83)(2.25,1.83)(2.25,2.17)
          (1.75,2.17)(1.75,1.83)
          \rput(2,2){$(2,1)$}
          \rput(2.68,2){\small \emph{transitive}}
          \psline[fillstyle=solid,fillcolor=lightgray](1.75,0.83)(2.25,0.83)(2.25,1.17)
          (1.75,1.17)(1.75,0.83)
          \rput(2,1){$(1,1)$}
          \rput(2.75,1){\small \emph{intransitive}}
%
%          \rput(-0.2,1.5){\footnotesize \emph{\textbf{extended preference}}}
%          \rput(-0.2,1.35){\tiny (it contains a linear order)}          
%          \psline[linestyle=dotted,linewidth=0.04,arrowsize=0.16]{->}(0.45,1.5)(1.93,1.5)
%
\end{pspicture}
\end{center}
\caption{\label{FIG:combinations_weak_Ferrers}
Ranking of desirability according to implications of $(m,n)$-Ferrers properties}
\end{figure}
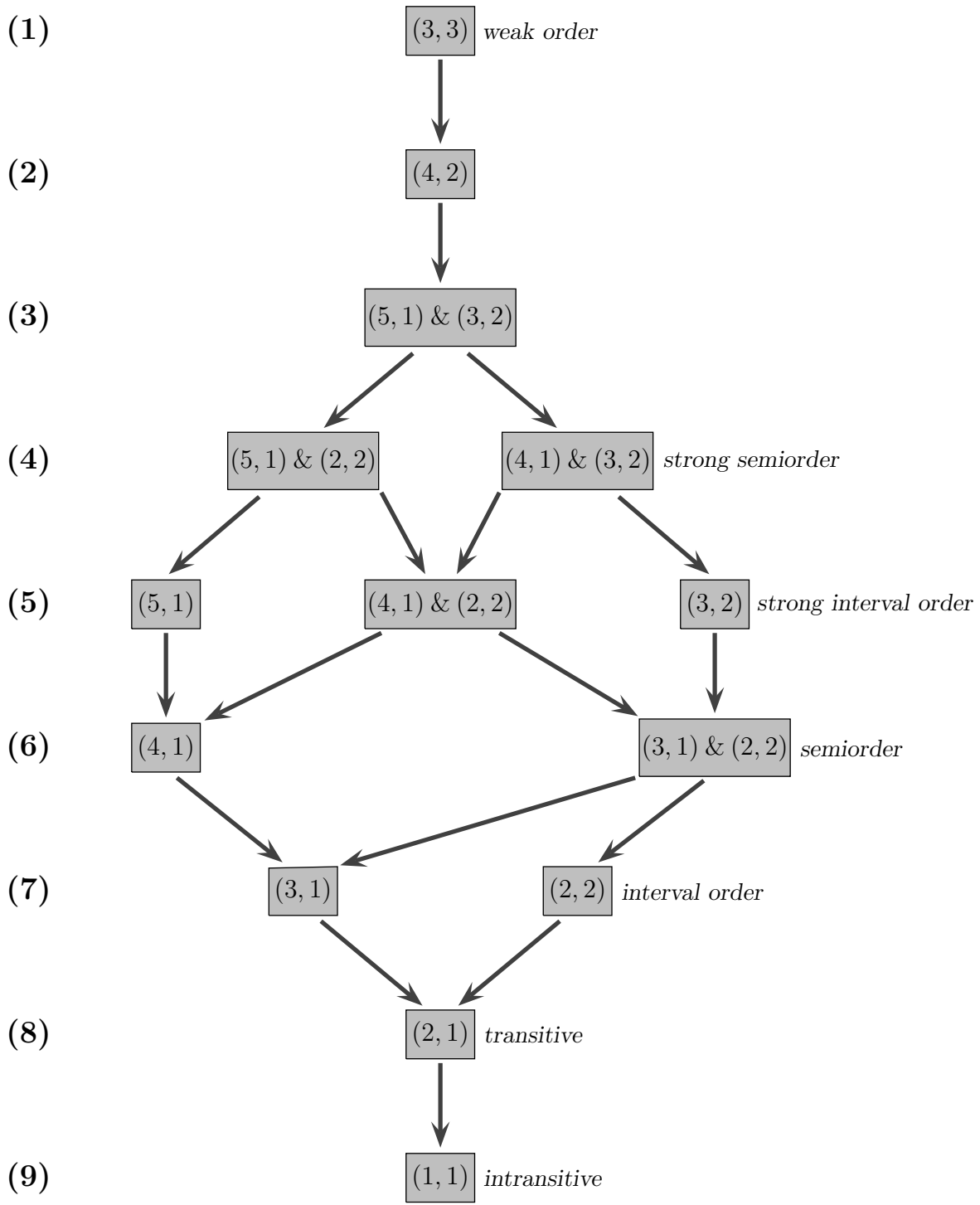

Consequently, all asymmetric and acyclic binary relation on a given set of alternatives can be partitioned according to a lattice structure, which is induced by the satisfaction of $(m,n)$-Ferrers properties.
This lattice is composed of 14 pairwise disjoint sets, which in turn can be arranged into 9 desirability classes according to their discrete degree of transitivity: see Figure~\ref{FIG:combinations_weak_Ferrers}.\footnote{This figure is a simple elaboration of Figure~6 in \cite{Gia2019}. See also \cite{Gia2014}, where the typical form of \textsl{strong semiorders} and \textsl{strong interval orders} is displayed in Figure~5.}  
For instance, the most desirable class is that of weak orders, whereas the least desirable class comprises all intransitive preferences. 
%Thus, denoted by $i = 1,\ldots , 9$ the nine desirability classes, and by $w$ a map assigning a weight to each class, we must have $w(i) \leqslant w(i +1)$.   
We can finally define a weighted variation of the degree of irrationality.  

%%%%%%%%   FIG. combinations of Ferrers properties %%%%%%%%%

%%%%%%%%%%%%%%%%%%%%%%%%%%%%%%%%%%%
%%%%%%%%%%%%%%%%%%%%%%%%%%%%%%%%%%%
%%%%%%%%%%%%%%%%%%%%%%%%%%%%%%%%%%%

\begin{definition}\label{DEF:distribution_metric} \rm
	A \textsl{feasible weighting map} is a function $w \colon \{1,2, \ldots,9\} \to (0,2)$, which assigns a positive weight to each desirability class in a way that 
	\begin{description}
		\item[\rm \textsl{(monotonicity)}] $i \leqslant j$ implies $w(i) \leqslant w(j)$ for all $i,j \in \{1,2,\ldots, 9\}$, and
		\item[\rm \textsl{(average property)}] $\sum_{i = 1}^9 \frac{w(i)}{9} \in \left[1-\varepsilon, 1 + \varepsilon \right]$ for some $0 \leqslant \varepsilon < 1$, 
	%	\item[\rm \textit{(average property)}] $\sum_{\succ \,\in \mathscr{A}(X)} \frac{w(\succ)}{\vert \mathscr{A}(X) \vert} \in \left[\frac{1}{2},\frac{3}{2}\right]$,
	\end{description}	
	where $\varepsilon$ is a discrimination threshold determined \textit{a priori} by the DM.\footnote{Here we do not dwell on the procedure to assess the discrimination threshold $\varepsilon$. In fact, the sole purpose of this section is to illustrate a simple variant of our approach.}
	Given a rationalizable choice $r$ over $X$, denote by $i_r$ the desirability class of its relation of revealed preference $\succ_r$.  
	Then, for any $c \in \choice(X)$, the \textsl{irrationality index of $c$ induced by $w$} is\vs
	\begin{equation*} \label{EQ:weighted_degree}
		\irr_{d_{\rat}}^{w} (c) :=\min\big\{w(i_r) \!\cdot\!d_{\rat}(c,r)  : r \in \choice_{\mathrm{rat}}(X) \big\}.
	\end{equation*}
\end{definition}

Definition~\ref{DEF:distribution_metric} can be motivated as follows. 
The property of monotonicity ensures that the weight of rational choices decreases as the level of transitivity of the corresponding revealed preference increases. 
As a consequence, if, for instance, a choice behavior is close to a highly desirable rational choice, then its degree of irrationality will be accordingly contracted. 
Furthermore, the average property guarantees that the average weight given to a rational choice behavior belongs to a close neighborhood of $1$ according to a threshold established by the DM.  
 
In the simplest case, all weights are the same, and the discrimination threshold is equal to $0$.
This implies that the weighting function $w$ assigns weight equal to $1$ to all asymmetric and acyclic binary relations over $X$. 
However, even in this very special case, it may happen that $\irr_\rat(c) \neq  \irr_\rat^w (c)$ for some choice $c$.
The reason is that the weighted variant of our approach only applies to decisive choice behaviors, and so the computation of the minimum distance from the benchmark of rationality may give different results. 

We conclude this section with an example, which showcases how a weighting procedure of rational choices yields a fine tuning of the results obtained in Example~\ref{EX:metric_new}. 

\begin{example} \rm \label{EX:stochastic_measure_of_deterministic_choice}
	Let $c_1,c_2,c_3,c_4$ be the four choice correspondences over $X = \{x,y,z,w\}$ defined in Example~\ref{EX:metric_new} (and further analyzed in Example~\ref{EX:metric_new_continued}). 
	%\footnote{Note that here we are restricting our attention to decisive choice behaviors, and therefore the irrationality degree computed in~\eqref{EQ:irrationality_of_choices_on_4_elements} may be different even without considering weighting procedures.} 
	On a four-element set, the phenomenology of $(m,n)$-Ferrers properties is quite poor, that is, many equivalence classes of the partition are empty. 
	In fact, it suffices to assign weights to the following classes: (1) weak orders, (6) semiorders, (7) interval orders and semitransitive relations, and (9) intransitive relations. 
	For the sake of illustration, first set $w(i) := 1 + 0.1 (i - 5)$ for $i = 1,\ldots , 9$. 
	%$w(6) := 1$, $w(7) := 1.1$, and $w(8) = w(9) := 1.4$. 
%	(For the other values of $w$, we may set them arbitrarily, as long as the monotonicity condition is verified.) %\footnote{We are aware of the arbitrariness of such an assignment of desirability levels. A robustness analysis may clarify the impact of choosing a different encoding.}  
	Clearly, $w$ is a feasible weighting map for any $0 \leqslant \varepsilon < 1$. 
	A computer-aided computation yields the following $d_\rat$-degrees of irrationality induced by $w$:\vs
	\begin{equation*} \label{EQ:irrationality_of_choices_on_4_elements_with_weight}
		\irr^{w}_{d_\rat} (c_1) = 0\,,\;\; \irr^{w}_{d_\rat} (c_2) = 5.4\,,\;\; \irr^{w}_{d_\rat}(c_3) = 12\,,\;\; \irr^{w}_{d_\rat} (c_4) = 13.2\,. 		
	\end{equation*}
	Now define $w' \colon \{1,2,\ldots , 9\} \to (0,2)$ by $w'(i) := 0.8$ for $1 \leqslant i \leqslant 4$, $w'(5) = 0.9$, and $w'(i) := 1.1$ for $6 \leqslant i \leqslant 9$. 
	Again, $w'$ is a feasible weighting map for any $0.2 \leqslant \varepsilon < 1$. 
	Now we get the $d_\rat$-degrees of irrationality induced by $w'$ become\vs
	\begin{equation*} \label{EQ:irrationality_of_choices_on_4_elements_with_weight}
		\irr^{w'}_{d_\rat} (c_1) = 0\,,\;\; \irr^{w'}_{d_\rat} (c_2) = 6.6\,,\;\; \irr^{w'}_{d_\rat}(c_3) = 16\,,\;\; \irr^{w'}_{d_\rat} (c_4) = 17.6\,. 		
	\end{equation*}
	A sensitivity analysis connected to the weighting procedure and the threshold of discrimination may provide further insight into the DM's preference system. 
\end{example}

%\textbf{\magenta Can we say something more formal in this section?}

%%%%%%%%%%%%%%%%%%%%%%%%%%%%%%%%%%%%%%%%%%%
%%%%%%%%%%%%%%%%%%%%%%%%%%%%%%%%%%%%%%%%%%%
%%%%%%%%%%%%%%%%%%%%%%%%%%%%%%%%%%%%%%%%%%%
%%%%%%%%%%%%%%%%%%%%%%%%%%%%%%%%%%%%%%%%%%%

\section{Measures of stochastic irrationality} \label{SECT:stochastic_setting}

In this last section we suggest how to adapt our approach to a stochastic environment. 
The underlying idea is to transform the search for a measure of irrationality into the formulation of a geometric problem (concerning polytopes). 

For simplicity, we shall only consider the case of stochastic choice functions,\footnote{Our approach can be extended to stochastic choice \textsl{correspondences}, too.} as defined below.  
 
\begin{definition}\label{DEF:SCF} \rm 
A \textsl{stochastic choice function} over $X$ is a map $p \colon X \times 2^X\setminus\{\es\} \to [0,1]$ such that for all $a \in X$ and $A \in 2^X \setminus \{\es\}$, the following conditions hold:\vs
\begin{itemize}
    \item $\sum_{a\in A}\, p(a,A)=1$,\vs
    \item $a \notin A$ implies $p(a,A)=0$.
\end{itemize}
We denote by $\choice^*(X)$ the family of all stochastic choice functions over $X$.
\end{definition}

As in the deterministic case, the first step in determining the degree of irrationality of a stochastic behavior consists of fixing a benchmark of rationality.

In their interesting approach, \cite{ApesteguiaBallester2015} essentially consider the \textit{finite} family of \textit{deterministic} rationalizable choices (which are in a one-to-one correspondence with linear orders) as the benchmark of rationality. 
Roughly speaking, the authors associate to a suitable stochastic choice behavior ---a \textsl{collection of observations}--- what they call a \textsl{swap index}, computed by using probabilities to weigh swaps in linear orders.
 
Our selection of the benchmark is instead an \textit{infinite} family of \textit{stochastic} choices, namely those that satisfy the following well-known model of rational behavior: 

\begin{definition}[\citealp{Block_Marschak1960}] \rm \label{DEF:RUM}
A stochastic choice function $p$ over $X$ satisfies the \textsl{random utility model} (for brevity, it is a \textsl{RUM} function) if there is a probability distribution $Pr$ on the set $\LO(X)$ of all linear orders over $X$ such that for each $A \in 2^X \setminus \{\es\}$ and $a \in A$,\vs 
$$
p(a,A) \;=\; Pr\big( \{\rhd \in \LO(X) : (\forall \,x \in A\setminus\{a\}) \;a \rhd x\} \big).\vs
$$
Hereafter, any RUM function will be called \textsl{rational}; accordingly, we shall denote by $\choice^*_\rat(X)$ the family of all RUM functions over $X$.   
\end{definition}

The selection of RUM as a prototype of stochastic rationality is statistically robust: see, among several related contributions, \cite{MarleyRegenwetter2017} for a review of random utility models, \cite{McCausland2019} for a direct Bayesian testing of RUM, and \citet[Section~8]{Davis-Stober2009} for an application to axiomatic measurement theory. 
%
%Due to the selection of RUM functions as a benchmark of rationality, an approach based on swaps ---in the style of \cite{ApesteguiaBallester2015}--- is unfeasible.

Now an attempt to fully adapt our deterministic approach to a stochastic setting poses salient challenges.   
In fact, we need an economically significant metric ---or, alternatively, a function that satisfies weaker properties, such as a `divergence'--- which enables us to discern different levels of irrationality for different types of stochastic choice behaviors. 
However, none of the metrics/divergences considered in the literature appears to be a good fit for our goal,\footnote{Some examples in Subsection~\ref{SUBSECT:examples} illustrate how different types of stochastic choice behaviors are not adequately distinguished by some well-known distances/divergences.} and it seems not simple to design new metrics that do the job.  

%
%Neither does it appear feasible to adapt our deterministic approach to a stochastic environment.
%Indeed, in the second and the third step in the evaluation of the irrationality of a stochastic choice behavior, we should\vs
%\begin{itemize}
%	\item[(2)] choose an economically significant metric ---or, alternatively, a function that satisfies weaker properties, such as a `divergence'--- which enables us to discern different levels of irrationality for different types of stochastic choice behaviors, and\vs 
%	\item[(3)] compute the minimum distance of irrational stochastic choices from the selected benchmark of rationality (that is, RUM functions) using the metric/divergence determined in step 2.
%\end{itemize}
%
%Effectively pursuing the above steps poses several problems. 
%In fact, none of the metrics/divergences considered in the literature appears to be a good fit for our goal, because they are unable to differentiate distinct behaviors. 
%(Some examples in Subsection~\ref{SUBSECT:examples} illustrate this fact.)   
%%Even more important, concerning step 3, in order to take the minimum distance from the benchmark of rationality, we need to compute all distances of a given behavior from (the infinite number of) RUM choices: this task is computationally unfeasible.

In view of the difficulties illustrated above, here we choose a different path to evaluate the level of irrationality of a stochastic choice behavior.  
Specifically, we take advantage of a known characterization of the RUM model to attach a vector with $\vert X \vert$-many components to each stochastic behavior: the higher the entries in the vector, the most irrational the choice behavior. 
Then, to compare irrationality levels, we use a permutation-invariant version of the classical Pareto ordering of these vectors, which arranges all irrational stochastic choices into a preordered set (ties and incomparability being allowed).
%The reminder of this section is devoted to present this novel measure of stochastic irrationality.  
%
%
%Specifically, first we use probability distributions to encode the desirability level of each benchmark of rationality, and then use the metric established in the previous section to determine the weighted distance from these rational choices. 
%
%Specifically, we shall define a rational measure of the irrationality of a stochastic choice function as the sum of all negative Block-Marschak polynomials.

As a preliminary step, we recall the known characterization of the RUM model.

\begin{definition}[\citealp{Block_Marschak1960,Falmagne1978}]\label{DEF:BM_inequalities} \rm 
Let $p$ be a stochastic choice function over $X$. 
For any $T \in 2^X \setminus \{\es\}$ and $a \in T$, define
$$
q_{a,T}:=\sum\limits_{T\subseteq U \subseteq X}	(-1)^{|U\setminus T|}p(a,U).
$$	
The $q_{a,T}$'s are the \textsl{Block-Marshak polynomials} (\textsl{BM polynomials}, for brevity)\footnote{`Polynomial' is the usual term, although $q_{a,T}$ is a linear expression in the $p(a,U)$'s} of $p$.
\end{definition}

\cite{Block_Marschak1960} show that having $q_{a,T}\ge 0$ for suitable menus $T \subseteq X$ is a necessary condition for having a RUM function.  
However, the general definition of the BM polynomials and the complete characterization of the random utility model came almost twenty years later:\footnote{See \cite{Fiorini2004} for an elegant and very short proof of this result, which involves M\"{o}bius inversion and network flow.} %\textbf{\magenta (Question: how does the work by Colonius, Fiorini, etc. directly relate to this? Need the reference to the work by JP and Saito.)}
 
\begin{theorem}[\citealp{Falmagne1978}]\label{THM:Falmagne}
A stochastic choice function is RUM if and only if all its Block-Marshak polynomials are nonnegative.
\end{theorem}

Theorem~\ref{THM:Falmagne} allows us to derive a measure of irrationality for stochastic choices. 
%\textbf{\magenta (In the final paper on stochastic choices, add considerations by Saito on capacities)}

\begin{definition} \rm \label{DEF:stochastic_irrationality_vector}
	Let $p$ be a stochastic choice function over $X = \{x_1,\ldots,x_n\}$, where $n \geqslant 2$.
	For each $x_i \in X$, let\vs\vs
	$$
	v_p(x_i) := 
% \max \left\{ - \sum \right\{q_{x_i,T} : q_{x_i,T} < 0 \right\}, 0 \right\}.
	\left\{
	\begin{array}{ll}
	    \left\vert \sum_{q_{x_i,T} < 0} q_{x_i,T}  \right\vert & \text{ if } q_{x_i,T} < 0 \text{ for some $T \in 2^X$,}\\
		0 & \text{ otherwise.}
	\end{array}\vs
	\right.\vs
	$$
	The $n$-tuple $v_p = \big(v_p(x_1),\ldots,v_p(x_n)\big) \in \R^n_+$ is the \textsl{negativity vector} of $p$.%\textbf{\magenta (Later on we may  consider the possibility to have a sum weighted by something depending on the cardinalities of the $T$'s)}
\end{definition}

Clearly, the larger the entries in the negativity vector, the more irrational the corresponding stochastic choice. 
By Definition~\ref{DEF:stochastic_irrationality_vector} and Theorem~\ref{THM:Falmagne}, all RUM functions ---and, in particular, all deterministic rationalizable choices--- have $(0,\ldots,0)$ as negativity vector.
For all non-RUM functions, the next definition establishes a way to compare their (strictly positive) irrationality.   

\begin{definition} \rm \label{DEF:preorder_of_irrationality}
	Denote by $\mathscr{S}(X)$ the family of all permutations of $X$.  
	Define a binary relation $\precsim^*$ over $\choice^*(X)$ as follows:\vs 
	$$
	p \precsim^* p' \quad \iff \quad (\exists \sigma \in \mathscr{S}(X))\: (\forall x \in X) \;  v_p(x) \leqslant v_{p'}(\sigma(x))\vs
	$$
	for any $p,p' \in \choice^*(X)$. 
	Then, we say that\vs
	\begin{itemize}
		\item $p$ and $p'$ are \textsl{equally irrational} if $p \sim^* p'$ (i.e., $p \precsim^* p'$ and $p' \precsim^* p$),\vs
		\item $p$ is \textsl{less irrational} than $p'$ if $p \prec^* p'$ (i.e., $p \precsim^* p'$ and $\neg(p' \precsim^* p)$), and\vs
		\item $p$ and $p'$ are \textsl{incomparably irrational} if $p \perp^* p'$ (i.e., $\neg(p \precsim^* p')$ and $\neg(p' \precsim^* p)$).
	\end{itemize}
The pair $\left(\choice^*(X),\precsim^* \right)$ is a preordered set,\footnote{Recall that a \textsl{preorder} is a reflexive and transitive (but possibly incomplete) binary relation.} having all RUM functions as a minimum.
\end{definition}
 
The next example presents two stochastic choice functions over a set of size four.
We shall compute all related BM polynomials and the two associated negativity vectors, to finally conclude that one function is more irrational than the other. 

\begin{example} \label{EX:negativity_vector} \rm
Set $X = \{x,y,z,w\}$.
Let $p_1$ be the stochastic choice function over $X$ defined in Table~\ref{TABLE:p1}.
For the sake of illustration, we explicitly compute the first two BM polynomials of $p_1$ associated to the item $x$:\vs
\begin{align*}
	q_{x,\{x\}} = & \quad p(x,\{x\}) - \big( p(x,\{x,y\}) + p(x,\{x,z\}) + p(x,\{x,w\}) \big) + \\
	& \quad \big( p(x,\{x,y,z\}) + p(x,\{x,y,w\}) + p(x,\{x,z,w\}) \big) - p(x,X) 	 \\
	= & \quad 1 - \big( 0.5 + 0.4 + 0.9 \big) + \big( 0.6 + 0.7 + 0.4 \big) - 0.4 \\ 
	= & \quad  0.5\,,\\
	q_{x,\{x,y\}} = & \quad p(x,\{x,y\}) - \big( p(x,\{x,y,z\}) + p(x,\{x,y,w\}) \big) + p(x,X) \\
	= & \quad 0.5 - \big( 0.6 + 0.7 \big) + 0.4 \\ 
	= & \quad - 0.4. 
\end{align*}
By Definition~\ref{DEF:stochastic_irrationality_vector}, summing all entries in the last four columns of Table~\ref{TABLE:p1}  yields the negativity vector of $p_1$, which is $v_{p_1} = (0.6,0.2,0.2,0.1)$.
	\begin{table}[h!] \footnotesize
	\begin{center}	
	\begin{tabular}{|l||||*{4}{c|}*{1}{|}*{4}{c|}}  
	\hline
	\backslashbox{\footnotesize menus}{\footnotesize items} 
	&\makebox[2em]{\footnotesize$x$} & \makebox[2em]{\footnotesize $y$} & \makebox[2em]{\footnotesize $z$} & \makebox[2em]{\footnotesize $w$} & \makebox[2em]{\footnotesize $q_{x,.}$} & \makebox[2em]{\footnotesize $q_{y,.}$} & \makebox[2em]{\footnotesize $q_{z,.}$} & \makebox[2em]{\footnotesize $q_{w,.}$} \\
	\hline\hline \hline \hline
	\footnotesize $\{x\}$ & $1$ & & & & $0.5$ & & & \\ 
	\hline
	\footnotesize $\{y\}$ & & $1$ & & & & {\red $-0.1$} & & \\ 
	\hline
	\footnotesize $\{z\}$ & & & $1$ & & & & $0.1$ & \\ 
	\hline
	\footnotesize $\{w\}$ & & & & $1$ & & & & $0.5$ \\ 
	\hline \hline
	\footnotesize $\{x,y\}$ & $0.5$ & $0.5$ & & & {\red $-0.4$} & $0.3$ & & \\ 
	\hline
	\footnotesize $\{x,z\}$ & $0.4$ & & $0.6$ & & {\red $-0.2$} & & $0.2$ & \\ 
	\hline
	\footnotesize $\{x,w\}$ & $0.9$ & & & $0.1$ & $0.2$ & & & $0$ \\ 
	\hline
	\footnotesize $\{y,z\}$ & & $0.5$ & $0.5$ & & & $0$ & $0.2$ & \\ 
	\hline
	\footnotesize $\{y,w\}$ & & $0.7$ & & $0.3$ & & $0.4$ & & $0.1$ \\ 
	\hline
	\footnotesize $\{z,w\}$ & & & $0.6$ & $0.4$ & & & {\red $-0.1$} & $0.3$ \\ 
	\hline \hline
	\footnotesize $\{x,y,z\}$ & $0.6$ & $0.3$ & $0.1$ & & $0.2$ & $0.1$ & {\red $-0.1$} & \\ 
	\hline
	\footnotesize $\{x,y,w\}$ & $0.7$ & $0.1$ & & $0.2$ & $0.3$ & {\red $-0.1$} & & $0$ \\ 
	\hline
	\footnotesize $\{x,z,w\}$ & $0.4$ & & $0.5$ & $0.1$ & $0$ & & $0.3$ & {\red $-0.1$} \\ 
	\hline
	\footnotesize $\{y,z,w\}$ & & $0.4$ & $0.4$ & $0.2$ & & $0.2$ & $0.2$ & $0$ \\ 
	\hline
	\footnotesize $\{x,y,z,w\}$ & $0.4$ & $0.2$ & $0.2$ & $0.2$ & $0.4$ & $0.2$ & $0.2$ & $0.2$ \\ 
	\hline 
	\end{tabular}\vs\vs\vs
	\end{center}
	\caption{\footnotesize The stochastic choice function $p_1$ and its BM polynomials: the entries in columns 1--4 give the probability that an item is chosen in a menu containing it, whereas the entries in columns 5--8 are the respective BM polynomials. (All empty entries stand for $0$.)} \label{TABLE:p1}
	\end{table}

A different stochastic choice function $p_2$ over $X$ is given in Table~\ref{TABLE:p2}.
	\begin{table}[h!] \footnotesize
	\begin{center}	
	\begin{tabular}{|l||||*{4}{c|}*{1}{|}*{4}{c|}}  
	\hline
	\backslashbox{\footnotesize menus}{\footnotesize items}
	&\makebox[2em]{\footnotesize $x$} & \makebox[2em]{\footnotesize $y$} & \makebox[2em]{\footnotesize $z$} & \makebox[2em]{\footnotesize $w$} & \makebox[2em]{\footnotesize $q_{x,.}$} & \makebox[2em]{\footnotesize $q_{y,.}$} & \makebox[2em]{\footnotesize $q_{z,.}$} & \makebox[2em]{\footnotesize $q_{w,.}$} \\
	\hline\hline \hline
	\footnotesize $\{x\}$ & $1$ & & & & $0.2$ & & & \\ 
	\hline
	\footnotesize $\{y\}$ & & $1$ & & & & $0.2$ & & \\ 
	\hline
	\footnotesize $\{z\}$ & & & $1$ & & & & $0.1$ & \\ 
	\hline
	\footnotesize $\{w\}$ & & & & $1$ & & & & $0.5$ \\ 
	\hline
	\hline
	\footnotesize $\{x,y\}$ & $0.6$ & $0.4$ & & & $0$ & $0.1$ & & \\ 
	\hline
	\footnotesize $\{x,z\}$ & $0.5$ & & $0.5$ & & $0$ & & $0.1$ & \\ 
	\hline
	\footnotesize $\{x,w\}$ & $0.8$ & & & $0.2$ & $0.2$ & & & $0$ \\ 
	\hline
	\footnotesize $\{y,z\}$ & & $0.4$ & $0.6$ & & & {\red $-0.1$} & $0.2$ & \\ 
	\hline 
	\footnotesize $\{y,w\}$ & & $0.7$ & & $0.3$ & & $0.3$ & & $0$ \\ 
	\hline
	\footnotesize $\{z,w\}$ & & & $0.6$ & $0.4$ & & & $0$ & $0.2$ \\ 
	\hline
	\hline
	\footnotesize $\{x,y,z\}$ & $0.5$ & $0.3$ & $0.2$ & & $0$ & $0.1$ & $0$ & \\ 
	\hline
	\footnotesize $\{x,y,w\}$ & $0.6$ & $0.2$ & & $0.2$ & $0.1$ & $0$ & & $0.1$ \\ 
	\hline
	\footnotesize $\{x,z,w\}$ & $0.5$ & & $0.4$ & $0.1$ & $0$ & & $0.2$ & $0$ \\ 
	\hline
	\footnotesize $\{y,z,w\}$ & & $0.4$ & $0.4$ & $0.2$ & & $0.2$ & $0.2$ & $0.1$ \\ 
	\hline
	\hline 
	\footnotesize $\{x,y,z,w\}$ & $0.5$ & $0.2$ & $0.2$ & $0.1$ & $0.5$ & $0.2$ & $0.2$ & $0.1$ \\ 
	\hline 
	\end{tabular}\vs\vs\vs
	\end{center}
	\caption{\footnotesize The stochastic choice function $p_2$ and its BM polynomials: all entries have the same meaning as in Table~\ref{TABLE:p1}.} \label{TABLE:p2}
	\end{table}
Note that $p_2$ provides a minimal counterexample to the fact that the property of \textsl{monotonicity}\footnote{A stochastic choice function $p$ over $X$ is \textsl{monotonic} (or \textsl{regular}) if for all $x \in X$ and $A,B \in 2^X$, $A \subseteq B$ implies $p_2(x,B) \leqslant p_2(x,A)$: see \cite{Block_Marschak1960}.} does not characterize the random utility model: in fact, $p_2$ is monotonic but not RUM. 

Since the negativity vector of $p_2$ is $v_{p_2} = (0,0.1,0,0)$, we get $v_{p_2}(a) \leqslant_{\Par} v_{p_1}(a)$ for all $a \in X$, and so we conclude that $p_2 \prec^* p_1$.  
\end{example}

As possibly expected, isomorphic stochastic choice functions ---in the sense clarified below--- are equally irrational. 

\begin{definition} \rm \label{DEF:isomorphic_stochastic_choices}
	Two stochastic choice functions $p,p'$ over $X$ are \textsl{isomorphic} is there is a permutation $\sigma \colon X \to X$ such that\vs
	$$
	p(x,A) = p'(\sigma(x), \sigma(A))\vs
	$$
	for all $x \in X$ and $A \in 2^X$. 
	The bijection $\sigma$ is called an \textsl{isomorphism} between $p$ and $p'$.  
\end{definition}

The next result shows that our measure of stochastic irrationality is independent of the names of alternatives. 

\begin{lemma}
	For any stochastic choices $p,p'$ over $X$, if $\sigma$ is an isomorphism between $p$ and $p'$, then $v_p(x) = v_{p'}(\sigma(x))$ for all $x \in X$.  
	Thus, isomorphic stochastic choice functions always have the same level of irrationality.  
\end{lemma}

\begin{proof}
	Observe that
	$$
	\sum\limits_{T\subseteq U \subseteq X}	\!\!\!(-1)^{|U\setminus T|}p(x,U) = \!\!\!\sum\limits_{T\subseteq U \subseteq X}\!\!\!	(-1)^{|\sigma(U)\setminus \sigma(T)|}p'(\sigma(x),\sigma(U))=\!\!\!\!\!\!\sum\limits_{\sigma(T)\subseteq U' \subseteq X}\!\!\!\!\!\!\!	(-1)^{|U'\setminus \sigma(T)|}p'(\sigma(x),U'),
	$$
	where the last equality is given by the fact that there is a one-to-one correspondences between the family of all menus $U$ containing $T$ and the family of all menus $U'$ containing $\sigma(T)$.
	We conclude that the BM polynomial $q_{x,T}$ of $p$ is equal to the BM polynomial $q_{\sigma(x),\sigma(T)}$ of $p'$. 
	The claim follows. 
	 \qed
\end{proof}

It is currently under study the implementation of a geometric approach (based on polytopes) to the measure of the irrationality of a stochastic choice behavior. 

%%%%%%%%%%%%%%%%%%%%%%%%%%%
%%%%%%%%%%%%%%%%%%%%%%%%%%%

\subsection{Some related literature} \label{SUBSECT:examples}

Here we review some existing metrics/divergences that apply to stochastic choices, and point out some possible drawbacks in detecting different levels of irrationality.   

\begin{definition}\rm \label{DEF:total_variation_distance} \rm
	Let $\delta \colon \choice^*(X) \times \choice^*(X) \to \R$ be the map defined by\vs 
	$$
	\delta(p,q)= \sup\vert p(a,A)-p'(a,A) \vert \vs 
	$$
	for all $p,p' \in \choice^*(X)$. 
	Then $\delta$ is a metric, called the \textit{total variation distance}.\footnote{This name originates from the process of considering all differences between two objects (stochastic functions, in this case) and taking either the sum or the supremum (the maximum, in this case).}
\end{definition}

The metric $\delta$ may not be a good fit for our purpose, as the next example shows.

\begin{example}\label{EX:total_variation_distance} \rm
	Define a stochastic choice function $p$ over $X=\{x,y,z\}$ by $p(a,A):=\frac{1}{\vert A \vert}$ for all $A \in 2^X \setminus \{\es\}$ and $a \in A$. 
	Clearly, $p$ is RUM function. 
	Next, we define two additional stochastic choice functions $p_1$ and $p_2$ over $X$ as follows:\vs 
	\begin{itemize}
		\item $p_1(a,A):=p(a,A)$ for all $A \in 2^X \setminus \{\es, X\}$ and $a \in A$,\vs
		\item $p_1(x,X):=0.6$, $p_1(x,X):=0.2$, and $p_1(z,X):=0.2$;\vs
		\item $p_2(x,\{x,y\}):=0.7$ and $p_2(y,\{x,y\}):=0.3$,\vs
		\item $p_2(x,\{x,z\}):=0.3$ and $p_2(z,\{x,z\}):=0.7$,\vs
		\item $p_2(y,\{y,z\}):=0.5$ and $p_2(z,\{y,z\}):=0.5$,\vs
		\item $p_2(x,X):=0.6$, $p_2(y,X):=0.3$, and $p_2(z,X):=0.1$.
	\end{itemize}
	It can be checked that $\delta(p,p_1)=\delta(p,p_2)= \frac{4}{15}$, %$0.6-\frac{1}{3}$, 
	that is, the metric $\delta$ puts $p_1$ and $p_2$ at the same total variation distance from the rational function $p$. 
	However, $v_{p_1} =(0.2,0,0)$ and $v_{p_2} = (0.3,0,0.1)$, and so $p_1 \prec^* p_2$ by Definition~\ref{DEF:preorder_of_irrationality}. 
\end{example}

Next, we consider a weaker type of distance, namely a `divergence', which only satisfies the non-negativity property \textsf{A0.1} of a metric, but not necessarily symmetry \textsf{A0.2} and the triangle inequality \textsf{A0.3}.

\begin{definition}[\citealp{KulLeb1951}]\label{DEF:Kullback_Leibler_divergence} \rm 
	Let $D_{\mathrm{KL}} \colon \choice^*(X) \times \choice^*(X) \to \R$ the function defined by\vs
	$$
	D_{\mathrm{KL}}(p \vert \vert p')= \sum_{(a,A)}p(a,A) \,\log\frac{p(a,A)}{p'(a,A)}\vs
	$$	
	for all $p,p' \in \choice^*(X)$. 
	%Moreover, let $\JSD(p \vert \vert q)=\frac{1}{2}D_{KL}(p \vert \vert m) + \frac{1}{2}D_{KL}(q \vert \vert m)$, where $m=\frac{1}{2}(p+q)$.
	%The function $\JSD$ is called the \textsl{Jensen-Shannon divergence}.
	The map $D$ is called the \textsl{Kullback-Leibler divergence}.
\end{definition}

\begin{example}\label{EX:Kullback_Leibler_divergence} \rm
	Let $X$, $p$, $p_1$, and $p_2$ be exactly as in Example~\ref{EX:total_variation_distance}.
	Define $p_3$ as follows:\vs
	\begin{itemize}
	    \item $p_3(x,\{x,y\})=0.7$ and $p_3(y,\{x,y\})=0.3$,\vs
		\item $p_3(x,\{x,z\})=0.3$ and $p_3(z,\{x,z\})=0.7$,\vs
		\item $p_3(y,\{y,z\})=0.7$ and  $p_3(z,\{y,z\})=0.3$,\vs
		\item $p_3(x,X)=\frac{1}{3}$, $p_3(y,X)=\frac{1}{3}$, and $p_3(z,X)=\frac{1}{3}$.
	\end{itemize}
	Note that the negativity vector of $p_3$ is $(0.03,0.03,0.03)$.  
	One can check that $D_{\mathrm{KL}}(p_1 \vert \vert p) < D_{\mathrm{KL}}(p_3 \vert \vert p) <D_{\mathrm{KL}}(p_2 \vert \vert p)$.
	On the other hand, according to Definition~\ref{DEF:preorder_of_irrationality}, we have $p_3 \perp^* p_1$ and $p_3 \perp^* p_2$. 
\end{example}

Examples~\ref{EX:total_variation_distance} and~\ref{EX:Kullback_Leibler_divergence} show that both the total variation distance and the Kullback-Leibler divergence may fail to capture some features of irrationality. 
Although one may argue that both examples only deal with one rational function ---possibly the most emblematic---, a similar pathology is still present when calculating distances from other rational functions.  
These issues suggest that Definition~\ref{DEF:preorder_of_irrationality} may provide a more adequate tool in assigning levels of irrationality to stochastic choices. 

%%%%%%%%%%%%%%%%%%%%%%%%%%%%%%%%%%%%%%%%%%%%%%
%%%%%%%%%%%%%%%%%%%%%%%%%%%%%%%%%%%%%%%%%%%%%%
%%%%%%%%%%%%%%%%%%%%%%%%%%%%%%%%%%%%%%%%%%%%%%
%%%%%%%%%%%%%%%%%%%%%%%%%%%%%%%%%%%%%%%%%%%%%%

%\subsection*{Statements and Declarations}
% 
% The authors have no relevant financial or non-financial interests to disclose. 
%Alfio Giarlotta acknowledges the support of ``Ministero dell'Istruzione, dell'Universit\`a e della Ricerca (MIUR) - PRIN 2017'', project \textit{Multiple Criteria Decision Analysis and Multiple Criteria Decision Theory}, grant 2017CY2NCA. 

%%%%%%%%%%%%%%%%%%%%%%%%%%%%%%%%%%%%%%%%%%%%%%
%%%%%%%%%%%%%%%%%%%%%%%%%%%%%%%%%%%%%%%%%%%%%%
%%%%%%%%%%%%%%%%%%%%%%%%%%%%%%%%%%%%%%%%%%%%%%
%%%%%%%%%%%%%%%%%%%%%%%%%%%%%%%%%%%%%%%%%%%%%%


\begin{thebibliography}{40}

\providecommand{\natexlab}[1]{#1}
\providecommand{\url}[1]{\texttt{#1}}
\expandafter\ifx\csname urlstyle\endcsname\relax
  \providecommand{\doi}[1]{doi: #1}\else
  \providecommand{\doi}{doi: \begingroup \urlstyle{rm}\Url}\fi
 \small
 
\bibitem[Aizerman and Aleskerov(1995)]{AizermanAleskerov1995}
\textsc{Aizerman, M., and Aleskerov, F.}, 1995. 
Theory of Choice. \textit{North Holland}.\vs
%  
\bibitem[Aleskerov and Monjardet(2002)]{AleMon2002} 
\textsc{Aleskerov, F., and Monjardet, B.}, 2002. 
Utility Maximization, Choice and Preference. \textit{Springer, Berlin}.\vs
%
\bibitem[Alabayrak and Aleskerov(2000)]{AlabayrakAleskerov2000}
\textsc{Albayrak, S.\,R., and Aleskerov, F.}, 2000.
Convexity of choice function sets. 
In: \textit{Bogazici University Research Paper} ISS/EC-2000-01.
%
\bibitem[Alcantud et al.(2022)]{AlcantudCantoneGiarlottaWatson2022}
\textsc{Alcantud, J.\,C.\,R., Cantone, D., Giarlotta, A., and Watson, S.}, 2022.
Rationalization of indecisive choice behavior by majoritarian ballots.
ArXiv:\:2210.16885 [econ.TH].
%
\bibitem[Ambrus and Rozen(2014)]{AmbrusRozen2014} 
\textsc{Ambrus, A., and Rozen, K.}, 2014. 
Rationalising choice with multi-self models. 
\textit{Economic Journal} 125(585): 1136--1156.
%
\bibitem[Apesteguia and Ballester(2015)]{ApesteguiaBallester2015}
{\textsc{Apesteguia, J., and Ballester, M.\,A.}, 2015.
A measure of rationality and welfare.
\textit{Journal of Political Economy} 123:\,1278--1310.}
%
\bibitem[Arrow(1950)]{Arrow1950} 
\textsc{Arrow, K.\,J.}, 1950. 
A difficulty in the concept of social welfare. 
\textit{Journal of Political Economy} 58: 328--346.
%
\bibitem[Arrow (1959)]{Arrow1959} 
\textsc{Arrow, K.\,J.}, 1959.
Rational choice functions and orderings. 
\textit{Economica} 26: 121--127.
%
\bibitem[Block and Marschak(1960)]{Block_Marschak1960}
\textsc{Block, H.\,D., and Marschak, J.}, 1960.
Random orderings and stochastic theories of responses.
In I.~Olkin, S.~Ghurye, H.~Hoeffding, W.~Madow, and H.~Mann, editors,
  \textit{Contributions to Probability and Statistics}, 97--132. Stanford
  University Press, Stanford.
%
%\bibitem[Bogart(1973)]{Bogart1973}
%\textsc{Bogart, K. P.}, 1973.
%Preference structures I: distances between transitive preference relations.
%\textit{Journal of Mathematical Sociology}, 3: 49--67.
%%
\bibitem[Cantone et al.(2016)]{CanGiaGreWat2016} 
\textsc{Cantone, D., Giarlotta, A., Greco, S., and Watson, S.}, 2016. 
$(m,n)$-rationalizability. 
\textit{Journal of Mathematical Psychology} 73: 12--27.
%%
\bibitem[Carpentiere et al.(2023)]{CarGiaWat2023}
\textsc{Carpentiere, D., Giarlotta, A., and Watson, S.}, 2023.
Corrigendum to: "A distance measure for choice functions'' [Social Choice and Welfare 30 (2008) 419--425].
Mimeo, University of Catania.
%\textit{Social Choice and Welfare} XXX, YYY.
%%
\bibitem[Chernoff(1954)]{Chernoff1954} 
\textsc{Chernoff, H.}, 1954. 
Rational selection of decision functions. 
\textit{Econometrica} 22: 422-443.
%
\bibitem[Costa-Gomes et al.(2022)]{Costa-GomezCuevaGerasimouTejiscak2022} 
\textsc{Costa-Gomes, M.\,A., Cueva, C., Gerasimou, G., and Teji\v s\v c\' ak, M.}, 2022. 
Choice, deferral, and consistency. 
\textit{Quantitative Economics} 13: 1297--1318.
%
\bibitem[Davis-Stober(2009)]{Davis-Stober2009}
\textsc{Davis-Stober, C.\,P.}, 2009.
Analysis of multinomial models under inequality constraints: Applications to measurement theory. 
\textit{Journal of Mathematical Psychology} 53: 1--13.
%
\bibitem[Falmagne(1978)]{Falmagne1978}
\textsc{Falmagne, J.-C.}, 1978.
A representation theorem for finite random scale systems.
\textit{Journal of Mathematical Psychology} 18: 52--72.
%
\bibitem[Fiorini(2004)]{Fiorini2004}
\textsc{Fiorni, S.}, 2004.
A short proof of a theorem of Falmagne. 
\textit{Journal of Mathematical Psychology} 48: 80--82.
%
\bibitem[Fishburn(1970)]{Fishburn1970} 
\textsc{Fishburn, P.\,C.}, 1970.
Intransitive indifference with unequal indifference intervals. 
\textit{Journal of Mathematical Psychology} 7: 144--149.
%
\bibitem[Fishburn(1985)]{Fishburn1985} 
\textsc{Fishburn, P.\,C.,} 1985.
\textit{Interval Orders and Interval Graphs}.
Wiley, New York.
%
\bibitem[Giarlotta(2014)]{Gia2014} 
\textsc{Giarlotta, A.}, 2014.
A genesis of interval orders and semiorders: Transitive \textsf{NaP}-preferences. 
\emph{Order} 31: 239--258.
%
\bibitem[Giarlotta(2019)]{Gia2019}
\textsc{Giarlotta, A.}, 2019. 
New trends in preference, utility, and choice: From a mono-approach to a multi-approach. 
In: M.\ Doumpos, J.\,R.\ Figueira, S.\ Greco, and C.\ Zopounidis (Eds.), \textit{New Perspectives in Multiple Criteria Decision Making}, Multiple Criteria Decision Making Series, Springer International Publishing, Cham, pp.\,3--80.
%
\bibitem[Giarlotta et al.(2022a)]{GiaPetWat2022a} 
\textsc{Giarlotta, A., Petralia, A., and Watson, S.,} 2022a. 
Bounded rationality is rare. 
\textit{Journal of Economic Theory} 204: 105509.
%
\bibitem[Giarlotta, et al.(2022b)]{GiarlottaPetraliaWatson2022b}
\textsc{Giarlotta, A., Petralia, A., and Watson, S.}, 2022b.
Context-sensitive rationality: Choice by salience. 
Available at SSRN: https://ssrn.com/abstract=4329891 or http://dx.doi.org/10.2139/ssrn.4329891.
%
\bibitem[Giarlotta and Watson(2014)]{GiaWat2014Ferrers} 
\textsc{Giarlotta, A., and Watson, S.}, 2014.
The pseudo-transitivity of preference relations: Strict and weak $(m, n)$-Ferrers properties.
\textit{Journal of Mathematical Psychology} 58: 45--54.
%
\bibitem[Giarlotta and Watson(2018)]{GiaWat2018} 
\textsc{Giarlotta, A., and Watson, S.,} 2018. 
Strict $(m, 1)$-Ferrers properties. 
\textit{Journal of Mathematical Psychology} 82: 84--96.
%
\bibitem[Kalai et al.(2002)]{KalaiRubinsteinSpiegler2002}
\textsc{Kalai, G., Rubinstein, A., and Spiegler, R.}, 2002.
Rationalizing choice functions by multiple rationales.
\textit{Econometrica} 70: 2481--2488.
%
\bibitem[Kemeny(1959)]{Kemeny1959}
\textsc{Kemeny, J. G.}, 1959.
Mathematics without numbers.
\textit{Daedalus} 88: 577--591.
%
\bibitem[Klamler(2008)]{Klamler2008}   
\textsc{Klamler, C.}, 2008. 
A distance measure for choice function,
\textit{Social Choice and Welfare} 30: 419--425.
%
\bibitem[Kreps(2013)]{Kreps2013}   
\textsc{Kreps, D.\,M.}, 2013. 
Microeconomic Foundation I. Choice and Competitive Markets.
\textit{Princeton University Press, Princeton}.
%
\bibitem[Kullback and Leibler(1951)]{KulLeb1951}
\textsc{Kullback, S., and Leibler, R. A.}, 1951.
On Information and Sufficiency.
\textit{Annals of Mathematical Statistics} 22: 79--86.
%
\bibitem[Luce(1956)]{Luce1956} 
\textsc{Luce, R.\,D.}, 1956.
Semiorders and a theory of utility discrimination.
\textit{Econometrica} 24: 178--191.
%
\bibitem[Manzini and Mariotti(2007)]{ManziniMariotti2007} 
\textsc{Manzini, P., and Mariotti, M.,} 2007. 
Sequentially rationalizable choice. 
\textit{American Economic Review} 97: 1824--1839.
%
\bibitem[Marley and Regenwetter(2017)]{MarleyRegenwetter2017}
\textsc{Marley, A.\,A.\,J., and Regenwetter, M.,} 2017. 
Choice, preference, and utility: probabilistic and deterministic representations. 
In: W.\,H. Batchelder, H. Colonius, E.\,N. Dzhafarov, and J. Myung (Eds.), 
\textit{New handbook of mathematical psychology: Vol.\:1}, Cambridge University Press, pp.\,374--453

\bibitem[Mas-Colell et al.(1995)]{Mas-ColellWhinstonGreen1995} 
\textsc{Mas-Colell, A., Whinston, M.\,D., and Green, J.\,R.}, 1995. 
\textit{Microeconomic Theory.}
Oxford University Press, New York.
%
\bibitem[Masatlioglu et al.(2012)]{MasatliogluNakajimaOzbay2012}
\textsc{Masatlioglu, Y., Nakajima, D., and Ozbay, E.\,Y.}, 2012.
Revealed attention. 
\textit{American Economic Review} 102: 2183--2205.
%
\bibitem[McCausland et al.(2019)]{McCausland2019}
\textsc{McCausland, W.\,J., Davis-Stober, C.\,P., Marley, A.\,J.\,J., Park, S., and Brown, N.}, 2019.
Testing the random utility hypothesis directly. 
\textit{The Economic Journal} 130(625): 183--207.
%
\bibitem[Nehring(1997)]{Nehring1997}
\textsc{Nehring, K.}, 1997. 
Rational choice and revealed preference without binariness. 
\textsl{Social Choice and Welfare} 14(3): 403--425.
%
\bibitem[Nishimura and Ok(2022)]{NishimuraOk2022}
\textsc{Nishimura, I., and Ok, E.\,A.}, 2022.
A class of dissimilarity semimetrics for preference relations. 
ArXiv:\:2203.04418 [math.CO].
%
\bibitem[Samuelson(1938)]{Samuelson1938} 
\textsc{Samuelson, P.}, 1938.
A note on the pure theory of consumers' behavior. 
\textit{Economica} 5: 61--71.
%
\bibitem[Sen(1971)]{Sen1971} 
\textsc{Sen, A.,} 1971.
Choice functions and revealed preferences. 
\textit{Review of Economic Studies} 38: 307--317.
%
\bibitem[Simon(1955)]{Simon1955}
\textsc{Simon, H.\,A.}, 1955.
A behavioral model of rational choice.
\textit{Quarterly Journal of Economics} 69: 99--118.
%
\end{thebibliography}
\end{document}